%% file: paper.tex
\documentclass[lettersize,journal]{IEEEtran}
\usepackage{amsmath,amsfonts}
\usepackage{array}
\usepackage[caption=false,font=scriptsize]{subfig}
\usepackage{textcomp}
\usepackage{stfloats}
\usepackage{url}
\usepackage{verbatim}
\usepackage{graphicx}
\usepackage{cite}
\usepackage{mathtools}
\usepackage{booktabs}
\usepackage{commath}
\usepackage{multirow}
\usepackage{amssymb}
\usepackage{siunitx}
\usepackage{amsthm}
\usepackage[ruled,linesnumbered]{algorithm2e}
\usepackage{epstopdf}

\usepackage{interval}
\usepackage{tikz,adjustbox}
\usepackage{circledsteps}
\usepackage{enumitem}
\usepackage{siunitx}
\usepackage{stackengine}
\usepackage{xcolor}
\usepackage[normalem]{ulem}
\newcommand\redsout{\bgroup\markoverwith{\textcolor{red}{\rule[0.5ex]{2pt}{1pt}}}\ULon}

\newcommand\demointro{122pt}
\newcommand\exdbuniform{124pt}
\newcommand\exdbnonuniform{124pt}
\newcommand\nonuniform{93pt}
\newcommand\exdnfastslow{122.5pt}
\newcommand\exdbevsnum{123pt}
\newcommand\exdbevsnuml{127pt}
\newcommand\exdbextreme{122.5pt}

\newcommand\exdbreg{204pt}
\newcommand\exdnsample{122.5pt}
\newcommand\exdbalphabeta{123pt}
\newcommand\secondproc{132pt}
\newcommand\secondprocc{95pt}
\newcommand\exdnfactor{123pt}
\newcommand\exdblimitations{123pt}

\newcommand\exiterations{82pt}
\newcommand\runtime{127.5pt}
\newcommand\demo{81pt}
\usepackage{xspace}
\makeatletter
\DeclareRobustCommand\onedot{\futurelet\@let@token\@onedot}
\def\@onedot{\ifx\@let@token.\else.\null\fi\xspace}
\def\eg{\emph{e.g}\onedot} 

\def\ie{\emph{i.e}\onedot}

\def\etal{\emph{et al}\onedot}
\makeatother
\usepackage[capitalize]{cleveref}
\crefname{section}{Sec.}{Secs.}
\Crefname{section}{Section}{Sections}
\Crefname{table}{Table}{Tables}
\crefname{table}{Tab.}{Tabs.}
\AtBeginEnvironment{appendices}{\crefalias{section}{appendix}}
\DeclareSIUnit\eps{EPS}
\DeclareSIUnit\fps{FPS}

\usepackage{fancyhdr}
\makeatletter
\let\ps@IEEEtitlepagestyle\ps@fancy
\makeatother
\begin{document}
\title{Neuromorphic Imaging with Joint Image Deblurring and Event Denoising}
\author{Pei Zhang,~\IEEEmembership{Graduate Student Member, IEEE}, Haosen Liu, Zhou Ge, Chutian Wang and Edmund Y. Lam,~\IEEEmembership{Fellow, IEEE}%
\thanks{This work was supported in part by the Research Grants Council of Hong Kong SAR (GRF 17201620, 17200321) and by ACCESS --- AI Chip Center for Emerging Smart Systems, sponsored by InnoHK funding, Hong Kong SAR.}%
\thanks{The authors are with the Department of Electrical and Electronic Engineering, The University of Hong Kong, Pokfulam, Hong Kong SAR, China (e-mail: \{zhangpei, hsliu, gezhou, ctwang, elam\}@eee.hku.hk). Edmund Y. Lam is also affiliated with ACCESS --- AI Chip Center for Emerging Smart Systems, Hong Kong Science Park, Hong Kong SAR, China.}%
\thanks{Corresponding author: Edmund Y. Lam.}}

\maketitle

\begin{abstract}
Neuromorphic imaging reacts to per-pixel brightness changes of a dynamic scene with high temporal precision and responds with asynchronous streaming events as a result. It also often supports a simultaneous output of an intensity image. Nevertheless, the raw events typically involve a large amount of noise due to the high sensitivity of the sensor, while capturing fast-moving objects at low frame rates results in blurry images. These deficiencies significantly degrade human observation and machine processing. Fortunately, the two information sources are inherently complementary --- events with microsecond-level temporal resolution, which are triggered by the edges of objects recorded in a latent sharp image, can supply rich motion details missing from the blurry one. In this work, we bring the two types of data together and introduce a simple yet effective unifying algorithm to jointly reconstruct blur-free images and noise-robust events in an iterative coarse-to-fine fashion. Specifically, an event-regularized prior offers precise high-frequency structures and dynamic features for blind deblurring, while image gradients serve as a kind of faithful supervision in regulating neuromorphic noise removal. Comprehensively evaluated on real and synthetic samples, such a synergy delivers superior reconstruction quality for both images with severe motion blur and raw event streams with a storm of noise, and also exhibits greater robustness to challenging realistic scenarios such as varying levels of illumination, contrast and motion magnitude. Meanwhile, it can be driven by much fewer events and holds a competitive edge at computational time overhead, rendering itself preferable as available computing resources are limited. Our solution gives impetus to the improvement of both sensing data and paves the way for highly accurate neuromorphic reasoning and analysis.
\end{abstract}

\input{./src/demo_intro}
\section{Introduction}\label{sec:introduction}
\IEEEPARstart{N}{euromorphic} cameras, equipped with a dynamic vision sensor (DVS), output asynchronous events within microseconds in response to pixel-level brightness changes of a dynamic scene, such that they are competent at recording fast motion with reduced blur~\cite{gallego2022event}. Some cameras that are also bundled with an active pixel sensor (APS) can capture events and intensity frames in parallel~\cite{taverni2018front}. Such a bio-inspired imaging has been bringing new applications in recent years~\cite{maqueda2018event,ge2022lens,li2022asynchronous,guo2023low}. However, two issues often arise in many scenarios:
\begin{enumerate}
    \item The DVS reacts to dynamic objects, posing challenges to the low frame-rate APS that produces blurry images when there is a relative movement between the camera and imaging scenes during the exposure time. Although the DVS suffers much less from motion blur, it presents a paradigm shift against the APS in information acquisition and loses low-frequency components by which human vision identifies fine-grained things.
    
    \item The DVS is susceptible to various sources of interference (\eg, sensor jitter and harsh illumination) and can return events with a lot of noise in response. Noise appears in the pixels with a lack of any active motion and substantially lowers processing accuracy.
\end{enumerate}

As such, the camera typically generates blurry images and noisy event streams during the imaging of highly dynamic scenes, as \cref{fig:demo_intro}~(a). It can be exacerbated for low illumination conditions, in which the noise level of the DVS becomes higher~\cite{lichtsteiner2008128} and the images of the APS are more blurry due to the increasing exposure time for image formation, leading to significantly degraded observations and analysis.

The key to solving these challenges is the interplay between asynchronous events and synchronous images, where the strengths of each can be leveraged to compensate for the deficiencies of the other. Events with microsecond temporal precision encode abundant motion information and can implicitly supply blind deblurring with a latent ground truth. On the other hand, while noise is irregular in space-time, informative events arise from the edges of a moving object that is clearly recorded in a sharp image, and the sharp image can thus offer accurate priors for event denoising. Accordingly, the challenging blind estimation of both images and events is simplified as the non-blind.

Fusing blurry images with events has delivered precise image improvements~\cite{scheerlinck2018continuous,pan2022high,wang2021asynchronous,zhang2022unifying}, and leveraging images as event noise filters has shown impressive results~\cite{wang2020joint}. Nevertheless, existing solutions only attend to one-sided enhancements and fail to recognize that the two kinds of data can collaborate with each other to jointly improve restoration quality. Aside from being much less practical, their processing is also heavily affected by the other side that is suffering from anomalous states. For example, the deblurring is subject to event noise and thus requires considerable events for support~\cite{scheerlinck2018continuous,pan2022high}, and the denoising can be malfunctioned by blurry images~\cite{wang2020joint}.

In this work, we introduce a unifying processing structure for the joint restoration of the blurry images and noisy events generated by a single neuromorphic camera, to simultaneously acquire blur-free images and noise-robust events (as shown in \cref{fig:demo_intro}). Specifically, an event-regularized prior is proposed for blind deblurring where the events are leveraged to supply the blurry image with auxiliary motion details. It depicts a blur-free high-frequency structure of the imaging scene and serves as a faithful ground truth to which the deblurred image features converge, rendering the improvement in the overall complexity and level of detail more effective and efficient. The recovered sharp image can afford gradient priors that faithfully supervise the discrimination between events and noise, due to events being triggered at moving object edges. With an iterative fashion, the two outputs can be progressively refined. Comprehensive assessments on real and synthetic datasets demonstrate that the synergy can deliver impressive restoration results, where the raw image suffers from severe motion blur, underexposure or overexposure, and the raw events are corrupted by substantial noise under strongly or weakly dynamic scenes. Meanwhile, it keeps a competitive edge at the event volume used and computational time consumption, making itself more practical for cases of limited computing resources. To fully evaluate the performance, we also collect a real neuromorphic dataset that consists of multiple pairs of blurry images and noisy event streams recorded by a DAVIS346 camera on a rich range of scenarios and settings.

This paper is structured as follows. Some pertinent studies are briefly reviewed in~\Cref{sec:related_work}. In~\Cref{sec:methodology}, we elucidate the proposed method that serves joint image and event reconstruction. \Cref{sec:experiments} elaborates experiment designs and the findings compared with other approaches. The paper ends with some concluding remarks in~\Cref{sec:conclusion}. We offer more supplementary analysis and discussions along with a few observations of the limitations of our work in Appendix.

\section{Related Work}\label{sec:related_work}
\subsection{Neuromorphic Imaging}
Neuromorphic cameras (or neuromorphic imaging), which have higher temporal resolution and higher dynamic range than conventional frame-based sensors, record local brightness changes in the form of sparse and asynchronous events~\cite{gallego2022event}. Such outstanding features make the imaging modality innately suitable for highly dynamic scenes under extreme illumination. Nevertheless, it generates time-sequence data streams that present a paradigm shift against frame-based imaging in visual information acquisition. To handle the streaming readout, some investigations tried to design alternative event-based manifestations (\eg, event frames~\cite{rebecq2017real} and graphs~\cite{li2021graph}), such that these new representations can inherit the advantages of events and can also be compatible with proven vision algorithms~\cite{baldwin2023time}. Accordingly, a variety of event-driven solutions have been developed for downstream applications, such as optical flow estimation~\cite{almatrafi2020distance,shiba22eccv}, motion prediction~\cite{maqueda2018event}, segmentation~\cite{jia2023event}, autofocusing~\cite{ge2023millisecond} and video frame interpolation~\cite{tulyakov2021time,tulyakov2022time}. In addition, increasing advances in processing technologies also give rise to a number of neuromorphic datasets, which are built either on simulators~\cite{hu2021v2e} or on real-scene capture~\cite{gehrig2021dsec}.

\subsection{Blind Image Deblurring}
Motion blur corrupts the images when there is a movement between the camera and the objects being captured. Blind image deblurring is an inverse process where we try to estimate a potential blur kernel and recover a latent sharp image~\cite{song2022dual}. There are two branches of research on image reconstruction, and one is the optimization only using a known blurry image, such as Pan~\etal~\cite{pan2017l_0} who proposed a revised $L_0$ regularization tailored for text images, Dong~\etal~\cite{dong2017blind} where an outlier-robust algorithm was developed to alleviate artifacts, and Pan~\etal~\cite{pan2019phase} in which the issue was addressed by effective image phase analysis. Recent studies are more in favor of learning-based methods that can bring greater robustness and higher accuracy from a mass of learned parameters~\cite{li2022learning}.

Since neuromorphic cameras can capture high-speed motion with microsecond precision, another branch tries to incorporate event features into image enhancements, taking advantages of the two information sources. Specifically, Scheerlinck~\etal~\cite{scheerlinck2018continuous} designed a novel complementary filter to bring frames into being a greater temporal resolution and dynamic range state. Pan~\etal~\cite{pan2019bringing,pan2022high} probed into an event-based double integral model to fuse blurry frames with the associated events and achieved impressive restoration. Wang~\etal~\cite{wang2021asynchronous} leveraged a kind of event-oriented augmentation whereby frames are enhanced with deblurring and temporal interpolation. Similarly, deep learning approaches are also applied to the neuromorphic field with compelling results in repairing videos with motion blur and enhancing visualization quality~\cite{jiang2020learning,lin2020learning,rebecq2021high,xu2021motion,zhang2022unifying,sun2022event,kim2022event}.

\subsection{Neuromorphic Noise Removal}
Sensitive neuromorphic cameras are highly vulnerable to various types of interference and can output noise as a result. This work is concerned with Background Activity (BA) noise. In bright scenes, the noise mainly results from junction leakage currents, while it is caused by thermal noise in dark scenes~\cite{guo2023low}. Besides, due to threshold mismatch issues, the cameras also have hot-pixels in which the noise is fired at an abnormally high rate at the same positions~\cite{hu2021v2e}. In relative to the noise behaving irregularly in space-time, events triggered by the edges of moving objects typically have strong continuity and correlation~\cite{zhang2023neuro}.

There have already been pioneering studies that achieved early progress on noise removal. Nearest neighbor-based filters built on local spatiotemporal correlation were first developed to suppress BA noise~\cite{delbruck2008frame}, and subsequent research presented substantial enhancements in memory optimization~\cite{khodamoradi2018n}. Offline approaches are being intensively studied due to their computing capacity and scalability, albeit they are potentially inappropriate for in-camera operations due to non-deterministic runtime. 
For example, event density~\cite{feng2020event,zhang2023neuro}, which suggests that events typically shape a denser cluster, and event probability~\cite{wu2020probabilistic}, where the denoising is modelled as an energy minimization process, can discriminate between events and noise. As artificial intelligence evolves, recent research also made use of deep neural networks to learn patterns on massive raw events for noise suppression~\cite{baldwin2020event,duan2021eventzoom}.

\section{Methodology}\label{sec:methodology}
\subsection{Neuromorphic Imaging Preliminary}
Once a logarithmic photocurrent change at a spatial position $\mathbf{x}$ exceeds a specific threshold at time $t$, neuromorphic cameras generate and encode an event $\mathbf{e}$, denoted by
\begin{equation}\label{eq:evs}
    \mathbf{e}_i = (\mathbf{x}, t_i, p_i),
\end{equation}
where $i$ indexes the event at a pixel $\mathbf{x} = (h, v)^{\mathsf{T}}$ consisting of two orthogonal directions $h$ and $v$. The symbol $p_i \in \{-1, +1\}$ indicates the sign of the change. As the time $t$ varies continuously, an event can also be modelled as
\begin{equation}\label{eq:evs_}
    \tilde{\mathbf{e}}_i(t) = p_i c \delta(t - t_i),
\end{equation}
where $\delta(t)$ is the Dirac delta function\footnote{In the following, we will alternatively use~\Cref{eq:evs} and~\Cref{eq:evs_} to describe an event.}, and $c$ is the temporal contrast threshold. Due to the function $\tilde{\mathbf{e}}_i(t)$ applying to every coordinate, we omit $\mathbf{x}$ in the above expression. As such, there is a mapping $f: \mathbf{e}_i \mapsto \tilde{\mathbf{e}}_i(t)$ from a spatiotemporal node to an impulse. Similarly, we define $\mathbf{E}$ for a batch of events
\begin{equation}\label{eq:event}
    \mathbf{E} = \{\mathbf{e}_i\}_{i=1:\infty} = \big\{(\mathbf{x}, t_i, p_i)\big\}_{i=1:\infty}.
\end{equation}
Then, a discrete event stream $\mathbf{E}$ turns into a continuous-time signal $f: \mathbf{E} \mapsto \tilde{\mathbf{E}}(t)$ that incorporates a train of impulses
\begin{equation}\label{eq:event_}
    \tilde{\mathbf{E}}(t) = \sum_{i=1}^{\infty} \tilde{\mathbf{e}}_i(t) = \sum_{i=1}^{\infty} p_i c \delta(t - t_i).
\end{equation}
We integrate $\tilde{\mathbf{E}}(t)$ over a time interval to obtain the quantized log intensity change $\mathbf{I}_\tau(t)$
\begin{equation}\label{eq:intensity}
    \mathbf{I}_\tau(t) = \int_{t-\tau}^{t} \tilde{\mathbf{E}}(\zeta)~\mathrm{d}\zeta = \int_{t-\tau}^{t} \sum_{i=1}^{\infty} p_i c \delta(\zeta - t_i)~\mathrm{d}\zeta,
\end{equation}
where a set of events within the interval $\tau$ are integrated, and then the increment for a short period $\Delta t$ is defined as 
\begin{equation}
    \Delta \mathbf{I}_\tau(t) = \mathbf{I}_\tau(t) - \mathbf{I}_\tau(t-\Delta t),
\end{equation}
which can be approximated by
\begin{equation}\label{eq:taylor}
    \Delta \mathbf{I}_\tau(t) \approx \frac{\partial}{\partial t}\mathbf{I}_\tau(t)\Delta t.
\end{equation}
Based on the brightness constancy assumption (\ie, optical flow constraint) for any short period $\Delta t$~\cite{gallego2015event}, we have
\begin{equation}\label{eq:constancy}
    \frac{\partial}{\partial t}\mathbf{I}_\tau(t) + \nabla_{\mathbf{x}}\mathbf{I}_\tau(t) \cdot \mathbf{v} = 0,
\end{equation}
where $\nabla_{\mathbf{x}} = (\partial_h, \partial_v)^{\mathsf{T}}$ symbolizes the differential operator, $\nabla_{\mathbf{x}}\mathbf{I}_\tau(t) = \left(\partial_h \mathbf{I}_\tau(t), \partial_v \mathbf{I}_\tau(t)\right)^{\mathsf{T}}$ is the spatial gradient, and $\mathbf{v} = (\frac{\mathrm{d}}{\mathrm{d} t} h, \frac{\mathrm{d}}{\mathrm{d} t} v)^{\mathsf{T}}$ denotes the motion field. We combine~\Cref{eq:taylor,eq:constancy} for a dot-product expression
\begin{equation}\label{eq:e_edge}
    \Delta \mathbf{I}_\tau(t) \approx - \nabla_{\mathbf{x}}\mathbf{I}_\tau(t) \cdot \mathbf{v}\Delta t,
\end{equation}
showing that events arise from the edges of an imaging object moving over a distance $\mathbf{v}\Delta t$. There is no event being triggered if the motion is parallel to the edges due to $\mathbf{v} \cdot \nabla_{\mathbf{x}}\mathbf{I}_\tau(t)  = 0$, and events are triggered at the maximum rate when the motion is perpendicular to the edges $\mathbf{v} \perp \nabla_{\mathbf{x}}\mathbf{I}_\tau(t)$.

\input{./src/demo}
\subsection{Joint Neuromorphic Reconstruction of Images and Events}\label{sec:id}
We formulate a uniformly blurry image $\mathbf{B}$ as the result of a convolution of a latent sharp image $\mathbf{S}$ with a spatially-invariant convolutional kernel $\mathbf{k}$ that possesses sufficient representational capability
\begin{equation}
    \mathbf{B} = \mathbf{S} \ast \mathbf{k} + r,
\end{equation}
where $r$ is the residual, and $\ast$ denotes the convolution operator. Blind image deblurring, which requires the simultaneous reconstruction of both $\mathbf{S}$ and $\mathbf{k}$ from available $\mathbf{B}$, is highly under-determined, since the same $\mathbf{B}$ can result from different pairs of $\mathbf{S}$ and $\mathbf{k}$. An unexpected solution is $\mathbf{S} = \mathbf{B}$ and $\mathbf{k}$ being a delta blur kernel. To achieve accurate restoration, the deblurring process can be modelled as
\begin{equation}\label{eq:deblur_process}
    \min_{\mathbf{S}, \mathbf{k}} \|\mathbf{S} \ast \mathbf{k} - \mathbf{B}\|_2^2 + \mathcal{R}_{\mathbf{S}} (\mathbf{S}) + \mathcal{R}_{\mathbf{k}} (\mathbf{k}),
\end{equation}
where the data-fidelity term expects the smallest difference between the estimated blurry image $\mathbf{S} \ast \mathbf{k}$ and the observation $\mathbf{B}$. Two regularizers $\mathcal{R}_{\mathbf{S}}$ and $\mathcal{R}_{\mathbf{k}}$ constrain $\mathbf{S}$ and $\mathbf{k}$, respectively. We decompose the deblurring into two subproblems and alternatively estimate $\mathbf{S}$ by solving
\begin{equation}\label{eq:estimate_s}
    \min_{\mathbf{S}} \|\mathbf{S} \ast \mathbf{k} - \mathbf{B}\|_2^2 + \mathcal{R}_{\mathbf{S}} (\mathbf{S})
\end{equation}
while keeping $\mathbf{k}$ fixed, and update $\mathbf{k}$ via
\begin{equation}\label{eq:estimate_k}
    \min_{\mathbf{k}} \|\mathbf{S} \ast \mathbf{k} - \mathbf{B}\|_2^2 +  \mathcal{R}_{\mathbf{k}} (\mathbf{k})
\end{equation}
by holding $\mathbf{S}$ constant. Through such iterative alternating minimization, $\mathbf{S}$ and $\mathbf{k}$ are progressively refined in each update, and the overall loss declines and converges to a minimum after iterations.
\input{./algorithms/alg}

As events record motion with microsecond precision and offer a blur-free depiction of high-frequency components (\eg, sharp edges and texture patterns) of a scene, they sufficiently represent the overall complexity and serve as a ground truth of the gradient space for estimating $\mathbf{S}$. In other words, events can be a prior to shape a sharp image in which the recorded object edges triggered the existing events. We thus introduce an event-based regularizer for supervising the estimation of $\mathbf{S}$, with
\begin{equation}
    \mathcal{R}_{\mathbf{S}} (\mathbf{S}) = \alpha \|\nabla_{\mathbf{x}} \mathbf{S} - \mathbf{I}_\tau(t)\|_2^2 + \beta \|\nabla_{\mathbf{x}} \mathbf{S}\|_0,
\end{equation}
where $\alpha$ and $\beta$ are penalty parameters. The first term imposes the spatial gradient of the deblurred image to converge toward the high-frequency structure described by the events that arise from an object in a latent sharp image. \cref{fig:demo} visualizes $\mathbf{S}$, $\nabla_{\mathbf{x}} \mathbf{S}$ and $\mathbf{I}_\tau(t)$ of the same scene respectively and also demonstrates the rationality of the regularizer. We also leverage an $L_0$ gradient prior~\cite{xu2011image} as the second constraint to suppress the undesirable artifacts, resulted from $\mathbf{I}_\tau(t)$ due to raw noisy events. Herein, we expect that the pixels of $\nabla_{\mathbf{x}} \mathbf{S}$, except for those on edges, do not have non-zero values. Then, we rewrite the objective function as
\begin{equation}\label{eq:estimate_s_re}
    \min_{\mathbf{S}} \|\mathbf{S} \ast \mathbf{k} - \mathbf{B}\|_2^2 + \alpha \|\nabla_{\mathbf{x}} \mathbf{S} - \mathbf{I}_\tau(t)\|_2^2 + \beta \|\nabla_{\mathbf{x}} \mathbf{S}\|_0.
\end{equation}
We handle the computationally intractable $L_0$-regularized term as earlier approaches~\cite{xu2011image,xu2013unnatural,pan2017l_0}, where \Cref{eq:estimate_s_re} is approximated by
\begin{equation}
    \min_{\mathbf{S}, \mathbf{z}} \|\mathbf{S} \ast \mathbf{k} - \mathbf{B}\|_2^2 + \alpha \|\nabla_{\mathbf{x}} \mathbf{S} - \mathbf{I}_\tau(t)\|_2^2 + \gamma \|\nabla_{\mathbf{x}} \mathbf{S} - \mathbf{z}\|_2^2 + \beta \|\mathbf{z}\|_0,
\end{equation}
where $\mathbf{z} = (z_h, z_v)^{\mathsf{T}}$ is an auxiliary variable associated with $\nabla_{\mathbf{x}} \mathbf{S} = (\partial_h \mathbf{S}, \partial_v \mathbf{S})^{\mathsf{T}}$, and $\gamma$ is a penalty weight. Likewise, it can be solved by alternating between evaluating $\mathbf{S}$ through
\begin{equation}\label{eq:estimate_s_sub}
    \min_{\mathbf{S}} \|\mathbf{S} \ast \mathbf{k} - \mathbf{B}\|_2^2 + \alpha \|\nabla_{\mathbf{x}} \mathbf{S} - \mathbf{I}_\tau(t)\|_2^2 + \gamma \|\nabla_{\mathbf{x}} \mathbf{S} - \mathbf{z}\|_2^2,
\end{equation}
and $\mathbf{z}$ via
\begin{equation}\label{eq:estimate_z}
    \min_{\mathbf{z}} \gamma \|\nabla_{\mathbf{x}} \mathbf{S} - \mathbf{z}\|_2^2 + \beta \|\mathbf{z}\|_0.
\end{equation}
Given the fixed $\mathbf{k}$ and $\mathbf{z}$, solving~\Cref{eq:estimate_s_sub} is a quadratic problem where the closed-form solution can be found by the Fast Fourier Transform (FFT)
\begin{equation}\label{eq:sol_s}
\mathbf{S} = \mathcal{F}^{-1} \text{\footnotesize$\left(\frac{\overline{\mathcal{F}}(\mathbf{k}) \mathcal{F}(\mathbf{B}) + \alpha \overline{\mathcal{F}}(\nabla_{\mathbf{x}})\mathcal{F}\left(\mathbf{I}_\tau(t)\right) + \gamma \hat{\mathcal{F}}(z_h, z_v)}{\overline{\mathcal{F}}(\mathbf{k}) \mathcal{F}(\mathbf{k}) + (\alpha + \gamma) \hat{\mathcal{F}}(\partial_h, \partial_v)}\right)$},
\end{equation}
where $\mathcal{F}, \mathcal{F}^{-1}$ and $\overline{\mathcal{F}}$ denote the FFT, inverse FFT and complex conjugate operator, respectively. The expression $\hat{\mathcal{F}}(\theta_0, \theta_1)$ with two arguments $\theta_0$ and $\theta_1$ is defined as
\begin{equation}
    \hat{\mathcal{F}}(\theta_0, \theta_1) = \overline{\mathcal{F}}(\partial_h)\mathcal{F}(\theta_0) + \overline{\mathcal{F}}(\partial_v)\mathcal{F}(\theta_1).
\end{equation}
Given $\mathbf{S}$, \Cref{eq:estimate_z} turns into element-wise minimization with hard thresholding~\cite{xu2011image}
\begin{equation}\label{eq:sol_z}
    \mathbf{z} = 
    \begin{cases} 
        0, & |\nabla_{\mathbf{x}} \mathbf{S}|^2 \leqslant \beta \gamma^{-1} \\ 
        \nabla_{\mathbf{x}} \mathbf{S}, & \text{otherwise}
    \end{cases}.
\end{equation}

Prior research evidences the estimation of $\mathbf{k}$ in the gradient domain to be superior with regard to the convergence rate and accuracy~\cite{cho2009fast}. As such, with $\mathbf{S}$ known, $\mathbf{k}$ of~\Cref{eq:estimate_k} can be updated by solving
\begin{equation}
    \min_{\mathbf{k}} \|\nabla_{\mathbf{x}} \mathbf{S} \ast \mathbf{k} - \nabla_{\mathbf{x}} \mathbf{B}\|_2^2 + \sigma \|\mathbf{k}\|_2^2,
\end{equation}
where $\sigma$ is a weight for the Gaussian regularizer $\mathcal{R}_{\mathbf{k}} (\mathbf{k}) = \|\mathbf{k}\|_2^2$. Similarly, this minimization can be solved by the FFT
\begin{equation}\label{eq:estimated_k}
    \mathbf{k} = \mathcal{F}^{-1} \left(\frac{\overline{\mathcal{F}}(\nabla_{\mathbf{x}} \mathbf{S}) \mathcal{F}(\nabla_{\mathbf{x}} \mathbf{B})}{\overline{\mathcal{F}}(\nabla_{\mathbf{x}} \mathbf{S}) \mathcal{F}(\nabla_{\mathbf{x}} \mathbf{S}) + \sigma}\right).
\end{equation}
We then set the negative elements of $\mathbf{k}$ to $0$ and normalize $\mathbf{k}$. 

BA noise is randomly distributed over the static background, while~\Cref{eq:e_edge} proves that events arise from moving edges. \Cref{eq:sol_s} recovers a sharp image $\mathbf{S}$ with accurate gradient information, whereby we channel neuromorphic noise removal under the supervision of the given prior $\nabla_{\mathbf{x}} \mathbf{S}$. We define a set of denoised events $\dot{\mathbf{E}} = \{\dot{\mathbf{e}}_i\}_{i=1:\infty} \subseteq \mathbf{E}$
\begin{equation}\label{eq:denoised_e}
        \dot{\mathbf{e}}_i = 
    \begin{cases} 
        \mathbf{e}_i, & g_j\sum_{i=1}^{\infty} \big|~p_i c \int_{-\infty}^{\infty}\delta(t - t_i)~\mathrm{d}t~\big| \neq 0\\ 
        \varnothing, & \text{otherwise}
    \end{cases},
\end{equation}
by which the inverse mapping $f^{-1}: \tilde{\mathbf{e}}_i(t) \mapsto \mathbf{e}_i $ is obtained. The mask $g_j \in \mathbf{g}$ simply follows
\begin{equation}\label{eq:sol_g}
        g_j = 
    \begin{cases} 
        0, & \nabla_{\mathbf{x}} \mathbf{S}_j \in \left(q-\omega, \; q+\omega\right)\\ 
        \nabla_{\mathbf{x}} \mathbf{S}_j, & \text{otherwise}
    \end{cases},
\end{equation}
where $\omega$ is the level of gradient supervision, $q$ is the pixel value of $\nabla_{\mathbf{x}} \mathbf{S}$ with the highest occurrences, and $\nabla_{\mathbf{x}} \mathbf{S}_j$, with $j$ indexing $\mathbf{x}$, is an element of $\nabla_{\mathbf{x}} \mathbf{S}$. \Cref{eq:denoised_e,eq:sol_g} describe that the useful signal should be in the position shared by both the raw events $\tilde{\mathbf{E}}(t)$ and the spatial gradient of a sharp image, coupling the two kinds of visual data in the denoising.

To prevent the events triggered by large motion that cannot be sufficiently represented by image edges from being filtered out, we harness the spatiotemporal correlation principle that events often temporally correlate with their spatial neighbors~\cite{khodamoradi2018n}, which can be mathematically expressed as
\begin{equation}\label{eq:neighbors}
    \mathbf{G}^{(i)}_{\mu, \nu} = \left\{\mathbf{e}_n \in \mathbf{E} \mid \mathcal{D}_{\mathbf{x}}(\mathbf{e}_n, \dot{\mathbf{e}}_i) \leqslant \mu, ~\mathcal{D}_t(\mathbf{e}_n, \dot{\mathbf{e}}_i) \leqslant \nu \right\},
\end{equation}
where an event $\mathbf{e}_n$ is regarded as one of the spatiotemporal neighbors $\mathbf{G}^{(i)}_{\mu, \nu}$ of $\dot{\mathbf{e}}_i$ if the two are adequately near in space-time. The functions $\mathcal{D}_{\mathbf{x}}$, $\mathcal{D}_t$ evaluate the spatial and temporal distance respectively, and $\mu$, $\nu$ are case-dependent thresholds that specify the boundary of the neighbors to be searched. As such, the expected signal comprises the events at the image edges and their space-time neighbors. 

Algorithm~\ref{alg} presents a dedicated workflow to bridge the two imaging modalities for jointly restoring the corresponding blur-free image and noise-robust events, given the input of a blurry image and noisy events from a neuromorphic camera. Such an iterative fashion enables coarse-to-fine reevaluations of $\mathbf{S}$ and $\dot{\mathbf{E}}$, while attenuating the noise interference from raw events (\ie, Line $2$) on $\mathbf{S}$ is realized by $\beta$ of~\Cref{eq:estimate_s_re}, where the pixels that are not on edges are penalized, and weakening that on $\dot{\mathbf{E}}$ is attained by $\omega$ of~\Cref{eq:sol_g}, by which the events in pixels with weaker intensity variations are filtered out. 

\subsection{Assumptions and Constraints}
The proposed approach is subject to the following assumptions and constraints:
\begin{itemize}
    \item There is always a latent sharp image in which the recorded object can effectively trigger events. This requires that the motion cannot be exactly parallel to object edges, and accordingly $\nabla_{\mathbf{x}}\mathbf{I}_\tau(t) \cdot \mathbf{v} \neq 0$ and  $\Delta \mathbf{I}_\tau(t) \neq 0$ in~\Cref{eq:e_edge}.

    \item In a monocular imaging system, we are concerned with the 2D projection of a 3D scene flow onto the sensor plane. The used camera lens with a low numerical aperture ensures that vertical motion and displacement on the $z$-axis are within the depth of field and thus negligible.

    \item Given the timestamp $t_b$ of a blurry image we are processing, $\tau$ of the event prior $\mathbf{I}_\tau(t)$ has $t_b = t - \frac{\tau}{2}$. That is, the sharp image at $t_b$ is blurred during the exposure time $[t_b-\frac{\tau}{2}, t_b+\frac{\tau}{2}]$.

    \item With the brightness constancy assumption holding for a short period $\Delta t$~\cite{gallego2015event}, we suppose $\mathbf{v}$ to be approximately invariant during $\Delta t$.

    \item A global contrast threshold $c$ is assumed to be unknown but constant during $\tau$ in~\Cref{eq:intensity}. Despite inevitably introducing a certain amount of error due to the variability nature of $c$, we observe that such a simple setting can still yield satisfactory results on real and simulated samples.
\end{itemize}

\input{./src/exdb_uniform}
\input{./src/uniform}
\input{./src/exdb_nonuniform}
\input{./src/nonuniform}

\section{Experiments}\label{sec:experiments}
\subsection{Blurry Images and Noisy Events Dataset}
We collect a real neuromorphic dataset made up of multiple pairs of blurry images and noisy events to evaluate our approach and other counterparts. The neuromorphic camera DAVIS346 MONO~\cite{taverni2018front}, with a spatial resolution of $346 \times 260$ pixels and a $120$ \si{\decibel} DVS dynamic range, is used to capture images and events in parallel from real dynamic scenes. We first record an object with the static camera for obtaining blur-free references, and then shake the camera during imaging such that the images suffer from uniform motion blur. After the similar acquisition of static recordings, we film an object moving at a high speed against a static background and thus acquire the results with non-uniform blur. To make images underexposed or overexposed, targets are placed under weak or strong illumination. Incandescent lighting with 60 \si{\Hz} serves as the source in the laboratory, and there is BA noise in raw events accordingly. Due to the illumination condition and the nature of the camera, our dataset does not include clean streaming events that can be used as the ground truth for exact quantitative analysis. This dataset also applies to the evaluation of traditional image reconstruction algorithms or pure event denoising approaches. More details are given in~\Cref{app:dataset}.

\subsection{Algorithm Parameters and Configurations}
We summarize the crucial parameters involved in our algorithm. The quantity $\tau$ is a period where there exists a log intensity change and determines the event volume accumulated during the interval. Its value is set by human perception and limited to milliseconds in our experiments. The penalty factors $\alpha$, $\beta$ and $\sigma$ regulate the weight of the event prior, $L_0$ gradient prior and Gaussian prior respectively. The parameter $\omega$ controls the level of gradient supervision given by an estimated sharp image, while the spatial and temporal thresholds $\mu$ and $\nu$ restrict the boundary of event neighbors to be searched. In the iterative processing, $l_{max}$ defines the maximum iteration, which is fixed to $5$ in our cases, and $\gamma_{max}$ sets a upper limit for the update of $\mathbf{S}$. The optimal values of the parameters vary with samples, and the resulting reconstructions should satisfy human observation. We analyze the influence of their settings on restoration quality in the following sections.

\subsection{Evaluations on Blind Image Deblurring}
Our solution and several state-of-the-art counterparts are assessed on sharpening a uniformly blurry image in~\cref{fig:exdb_uniform,fig:uniform}, and on enhancing the clarity of an image with non-uniform blur in~\cref{fig:exdb_nonuniform,fig:nonuniform}. Clear images of a scene are captured and serve as the (pseudo) ground truth, becoming a reference for the structures and shapes of the imaging scene to visually evaluate the reconstructions of a real blurry image. 

When recovering the sharp state of the blurry image with uniform motion, the image-based method, \cref{fig:exdb_uniform}~(e), suffers from serious ringing artifacts, whereas the event-based ones, \cref{fig:exdb_uniform}~(f)--(h), bring better clarity in comparison. However, they either leave blurry residues or contain noticeable gray flecks (caused by event noise). With the precise motion capture as a prior, our algorithm shapes a clearer image and visually outperforms all the competing counterparts by large margins. \cref{fig:uniform} explores whether the events from a real weakly dynamic scene~\cite{mueggler2017event} conduce to the removal of a synthetic strong blur, meanwhile simulating a realistic instance that capturing low-contrast targets under weak lighting often brings extremely blurry images yet very few events triggered. A synthetic kernel is used to blur a clear image, as~\cref{fig:uniform}~(a) and~(b). The event volume of the real scene falls far short of the quantity such a strong level of motion might generate, which makes the deblurring quite demanding. While the competitors underperform, our solution can still obtain a result with a satisfactory level of sharpness.

It becomes much more challenging to find the sharp shape of a non-uniform blurry image where imaging objects are moving against static scenes, such that there is inconsistent blur in the foreground and background. Our uniform deblurring method can be extended to non-uniform blur processing, and the explanations are given in~\Cref{app:nudb}. \cref{fig:exdb_nonuniform}~(b) highlights two positions with different blur states, and only one of them has a significant movement as~\cref{fig:exdb_nonuniform}~(c) reveals. Compared with the image-based method, the event-based ones obtain a higher level of clarity since events can precisely locate active pixels, and ours brings sharper visualization of texts and reduced artifacts on the pixels of the dynamic foreground. \cref{fig:nonuniform} researches on restoring a blurry image with areas of varying contrast to a sharp state~\cite{wang2020joint}, with~(b) highlighting the zone. As in~\cref{fig:nonuniform}~(c), the event quantity is significantly lower in the low-contrast pixels than those in the high-contrast ones due to weaker brightness changes that the camera hardly perceives, leading to nonuniform distribution of events in space-time and insufficient high-frequency priors. The results show that our method is more robust to this challenging scenario and can accurately recover the sharp shapes, lines and edges especially in the low-contrast areas.

\input{./src/exdb_extreme}
\input{./src/exdb_gt_tab}

Quantitative evaluations are conducted based on the DAVIS 240C Datasets~\cite{mueggler2017event}, where we leverage the Mean Squared Error (MSE) to measure the difference in pixels, Perceptual Loss (LPIPS with scaling) to evaluate the variation in deep features~\cite{zhang2018perceptual} and Structural Similarity Index Measure (SSIM) to assess the resemblance in structural details. An effective algorithm is expected to earn a low value of MSE, LPIPS, and a high value of SSIM. \Cref{tab:exdb_gt_tab} presents the comparisons, in which the type \texttt{Image} includes image-based methods, \texttt{Event} represents the event modality only and \texttt{Fusion} stands for the synergy of the two modalities. As it shows, our solution delivers the best results in all the metrics, indicating that our reconstructions and the original data have fewer differences in pixel values, deep features and structural information. Specifically, compared with the event-based counterparts, ours achieves nearly $50\%$ of the LPIPS of Scheerlinck~\etal's and surpasses the SSIM of Pan~\etal's by over $0.1$. The learning-based approaches~\cite{rebecq2021high,sun2022event} have comparatively lower LPIPS and higher SSIM values, while ours still has slight advantages over them in terms of the MSE and LPIPS.

\subsection{Event-based Reconstructions for Challenging Scenarios}
Events feature high temporal precision and high dynamic range, enabling clear recordings of fast-motion and the details almost missing from underexposed or overexposed images. \cref{fig:exdb_extreme} demonstrates four challenging scenarios --- a text object captured by a shaking camera under low illumination in which the blurry, underexposed image almost loses dark contents; a low-lighting pedestrian scene where the raw image is severely blurry, delayed and overexposed in some pixels due to a large period of exposure time necessary for image formation~\cite{scheerlinck2018continuous}; recording a box with letters on it under strong illumination, which leads to the overexposed image to lose shapes and textures; shooting a high-speed badminton with rich textures under sufficient lighting where the image still undergoes heavy motion blur. Due to incomplete information in the raw images, image-based methods cannot function properly to yield reasonable results. When all the approaches are supported by the same event quantity, our images have higher clarity of recovered details and look more realistic.

\input{./src/exdn_sample}

\input{./src/exdn_gt_tab}
\subsection{Evaluations on Neuromorphic Noise Removal}
The DVS is vulnerable when illumination is not constant and responds with a great amount of BA noise accordingly. \cref{fig:exdn_sample} shows the comparisons on neuromorphic noise removal, and (a) spotlights the challenging regions involving both events and noise. While the competitors either fail to suppress noise or erase lots of informative events, our method features more accurate discrimination.

Quantitative denoising analysis is performed on the samples \texttt{outdoors\_running} (\texttt{Running}), \texttt{shapes\_rotation} (\texttt{Shapes}) of the DAVIS 240C Datasets~\cite{mueggler2017event}, where each sample has synthetic random noise with the amount of $50\%$ of the raw event volume. Instead of the Signal-to-Noise Ratio that is biased when the aggressive denoising achieves a high score by removing noise along with a significant number of events, we consider the denoising as a binary classification task, in which a filter classifies an event as signal (positive class) or noise (negative class). The statistics --- the True Positive Rate (TPR), False Positive Rate (FPR), Precision (PPV) and Accuracy (ACC), are measures of the classifier performance. \Cref{tab:exdn_gt_tab} presents the comparisons, and the Nearest Neighbor Filter (NN-Filter) --- the prototype accepting the events with strong spatiotemporal correlation~\cite{czech2016evaluating}, acts as a baseline. Our approach correctly classifies signal and achieves the highest TPR, and then obtains the second best FPR and PPV behind Feng~\etal's. However, combining the low value of TPR of Feng~\etal's, we infer that it has good noise removal but also discards lots of true events. As such, neither one of TPR, FPR and PPV alone can sufficiently represent the performance. The ACC measures how accurate a method is to classify both signal and noise. Our algorithm acquires the highest ACC and thus features better denoising capabilities over the counterparts.

\input{./src/exdn_fast_slow}
\subsection{Denoising for Strongly and Weakly Dynamic Scenes}
Actively moving objects often trigger a considerable volume of events, whereas the denoising becomes more challenging when objects hardly move, such that the events have weaker spatiotemporal correlation and account for a lower proportion in number. We study the reactions of existing algorithms to these two cases in~\cref{fig:exdn_fast_slow}. In the raw data, we might visually distinguish events and noise for the fast-moving object based on pixel density but fail in the slow-moving one. The examined methods achieve comparable results in the first case, while when no configuration changes of each one for the slow-moving, the counterparts either leave visible noise residues or barely retain events. By comparison, our gradient-driven mechanism is less sensitive to such variations in event attributes and distribution, and features more reliable processing for the cases with a sharp drop in the event quantity.

\input{./src/exdb_evsnum}
\subsection{Analysis of Event Quantity for Image Deblurring}\label{sec:event_quantity}
The number of events used for the deblurring dominates reconstructed image quality. While increasing the quantity often supplies richer dynamic features and typically leads to better results, it also causes more processing latency. We expect good restoration built on the supplemental information from a small event volume. In~\cref{fig:exdb_evsnum}~(a) and~(b), we visually evaluate the deblurred images that are based on varying numbers of events. Our method recovers clearer images with the aid of very few events, and the counterparts still leave severe motion blur.

\cref{fig:exdb_evsnum}~(c) gives quantitative analysis on synthetic samples. Our method achieves much more outstanding scores based on a small event quantity collected within $40$ \si{\ms}. We note that our performance degrades with the increase of the event accumulation time. A potential reason could be that the excessive events used cover multiple scenes whose redundant information is irrelevant to the current latent sharp state. Similar issues can also be found in the competing methods, and the optimum of each relates to how events are represented in processing. The above evaluations show that our approach can be driven by much fewer events and has higher event utilization efficiency, which is preferable when available resources are limited. 

\input{./src/exdb_alpha_beta}
\input{./src/prior_quality}
\subsection{Analysis of the Regularizers for Deblurring}
As in~\Cref{eq:estimate_s_re}, the deblurring operation is constrained by an event-based regularizer and an $L_0$ term. \cref{fig:exdb_alpha_beta} analyzes how the weight $\alpha$, with a fixed $\beta$, affects the processed images. When $\alpha=0$, and the $L_0$ term becomes the only constraint, the blurry image is deficiently sharpened and slightly smoothed, indicating that the $L_0$ facilitates smoothness (or noise suppression) in our cases. As the event-based regularization receives a proper weight ($\alpha=0.24$) and functions well, the blur-free image can be faithfully reconstructed. Increasing the value of $\alpha$ enforces the image to approach the coarse event intensity in shades of gray and shapes, and to involve more artifacts resulting from event noise accordingly. \Cref{app:l0rd} shows the effect of the $L_0$ term. \cref{fig:exdb_alpha_beta}~(e) offers mensurable evaluations of how the constraints impact blur kernel estimation, and higher similarity generally reflects more accurate deblurring. When $\beta$ is fixed, increasing the value of $\alpha$ yields gradually steady performance, whereas the cases of $\beta$ exhibit an almost opposite tendency. Visual and quantitative assessments show that the proposed event prior supplies a precise description of the shapes, structures and positions of moving objects in a latent sharp image and thus dominates rough deblurred results, and the $L_0$ constraint works on auxiliary calibrations.

In addition, we investigate how the quality of the event prior impacts image deblurring. As in~\Cref{eq:intensity}, $\mathbf{I}_\tau(t)$ is obtained by integrating a set of events (modelled as a train of impulses) over a time interval $\tau$. Given a stream, the quality of $\mathbf{I}_\tau(t)$ is thus constrained by $\tau$. \cref{fig:prior_quality} presents the results based on the supplemental information given by $\mathbf{I}_\tau(t)$ of different $\tau$, where (a) is an underexposed blurry image with two areas of concern. A small interval $\tau = 2$ \si{\ms} leads to insufficient events collected and accordingly the limited expressive power of the prior, resulting in the deblurred image with vague details. \cref{fig:prior_quality}~(c) shows the best result as $\tau = 6$ \si{\ms} brings $\mathbf{I}_\tau(t)$ a good quality. When using an excessive number of events ($\tau = 16$ \si{\ms}) in which multiple scenes involved make the event prior degraded, the processed image still suffers from blur artifacts and also loses textures in the low-contrast area. The above findings are consistent with the quantitative analysis in~\Cref{sec:event_quantity}. 

\input{./src/exdn_factor}
\subsection{Analysis of Gradient Supervision for Event Denoising}
\Cref{eq:sol_g} exploits a factor $\omega$ to fine-tune the magnitude of gradient supervision imposed by a sharp image. \cref{fig:exdn_factor} studies its impact on denoised results. Setting a large value of $\omega$ suppresses sparse events and retains only the most active (densest) ones, showing that increasing its value is equivalent to enhancing denoising strength by raising the filter threshold. Valuable information might be lost although sufficient denoising can be obtained. The optimum $\omega$ varies with cases and seeks a trade-off --- a detector on autonomous vehicles might require a low $\omega$ to avoid dropping events triggered by small obstacles, whereas a data collector in optical systems might prefer a high value for capturing noise-robust data.

\subsection{Investigations on Computational Time Overhead}
Despite two kinds of visual data restoration integrated in one unifying structure, the computational cost each one takes can be analyzed individually by setting proper checkpoints. \cref{fig:runtime}~(a) compares runtime consumed for image deblurring and exhibits observable advantages of the event-based methods over the image-based one (Bai's~\cite{bai2018graph}). The time overhead taken by the former rises with the event volume. Our approach holds a slight competitive edge among the counterparts when a small event quantity is used (\eg, within $20$ \si{\ms}), and better scores can still be achieved as shown in~\cref{fig:exdb_evsnum}~(c). \cref{fig:runtime}~(b) explores how runtime changes with an increasing number of raw events to be denoised, where our curve with a relatively gentle rise shows that the solution is preferable for a system with strict latency constraints. The above observations reveal that the proposed joint algorithm can reconstruct both the deblurred image and denoised events within a tolerable delay.
\input{./src/runtime}

\section{Conclusion}\label{sec:conclusion}
When capturing dynamic scenes, neuromorphic cameras generate blurry images and noisy events due to the nature of the sensors. This work bridges the two data sources to reconstruct blur-free images and noise-robust events in parallel. Our simple yet effective solution shows more satisfactory results compared with other methods and manifests greater robustness to several challenging cases, paving the way for accurate high-level neuromorphic analysis. Nevertheless, as it relies on event information, we fail to faithfully recover a sharp image whose details are missing from the blurry one and are not recorded by the events. Limitations of our algorithm are discussed in~\Cref{app:limit}.

\input{./src/exdb_nonuniform_app}

\begin{appendices}
\input{./src/dataset_tab}
\section{Blurry Images and Noisy Events Dataset}\label{app:dataset}
The motivation of this data collection work lies in that there is an urgent need for neuromorphic datasets that include pairs of blurry images and noisy events. Our dataset covers a rich range of scenes and settings, involving uniform and non-uniform motion in weak and strong lighting, as well as challenging scenarios with increased event noise. \Cref{tab:dataset} provides more details about each sample. For different samples, the count of events ranges from $3 \times 10^5$ \si{\eps} (events per second) to $8 \times 10^6$ \si{\eps}, and the images are recorded at a frame rate from $1$ \si{\fps} (frames per second) to $25$ \si{\fps}.

\section{Extension to Non-Uniform Deblurring}\label{app:nudb}
We extend uniform deblurring to handle non-uniform motion blur, where a blurry image is split into multiple patches associated with different blur kernels. As such, a non-uniform blurry image $\mathbf{B}$ can be modelled as
\begin{equation}
    \mathbf{B} = \sum_{i = 1}^\infty \mathbf{B}_i = \sum_{i = 1}^\infty \mathbf{S}_i \ast \mathbf{k}_i,
\end{equation}
where the $i$th patch $\mathbf{S}_i$ of an image $\mathbf{S}$ is blurred by the corresponding kernel $\mathbf{k}_i$ to obtain the blurry patch $\mathbf{B}_i$. We can use our uniform deblurring to sharpen blurry patches and then acquire a complete sharp image by
\begin{equation}
    \mathbf{S} = \sum_{i = 1}^\infty \mathbf{S}_i.
\end{equation}
This operation builds upon the assumption that there is uniform and consistent motion over a sufficiently small area. The smaller the patch, the more accurate the estimation and approximation of the blur kernel, and therefore the better the deblurring quality.

As our algorithm incorporates event information, and events only record the foreground with significant motion, we explore the use of uniform deblurring for non-uniform blurry images, as shown in~\cref{fig:exdb_nonuniform_app}. The image-based competitor fails to deblur the image by uniform deblurring, since it has multiple conflicting kernels and cannot decide the correct one to estimate the sharp image. It then brings visible improvements by patch-by-patch processing. By contrast, our method achieves comparable results by the two treatments, although there are slight blur artifacts in the static background of our uniformly deblurred image. The study proves that event-based solutions have more precise positioning of the dynamic foreground and better robustness to various motion patterns.

\section{Analysis of the $L_0$ Regularizer for Deblurring}\label{app:l0rd}
\cref{fig:exdb_beta} offers more information about how the $L_0$ regularization affects deblurred images. When $\beta = 0.004$ and the event-based constraint still dominates, we have a clear, sharp result that also contains observable artifacts (\ie, white flecks). The artifacts are then suppressed when the $L_0$ term prevails with a high value of $\beta = 0.256$, but the texture and structure details recorded in the events are also smoothed and erased, and the associated blur kernel is distorted.
\input{./src/exdb_beta}
\input{./src/iterations}
\input{./src/second_proc}
\section{Visualization of the Iterative Reconstruction}\label{app:iterations}
As Algorithm~\ref{alg} iteratively evaluates the latent sharp image $\mathbf{S}$ and the denoised events $\dot{\mathbf{E}}$ in a coarse-to-fine fashion, we visualize the results of each iteration in~\cref{fig:iterations}. As shown, the raw image is significantly degraded by severe motion blur, and the raw events also contain a storm of noise. The joint improvement undergoes $l_{max}=5$ iterations, with each one showing increasing enhancements. The gradually optimized sharp image with clearer shapes and textures provides gradient supervision for event search from the raw noisy data, resulting in an increasing quantity of events being retrieved. We also notice that the general structure can already be accurately estimated in the first iteration, indicating that the event prior sufficiently supplies the deblurring with a precise description of high-frequency features of a dynamic scene, contributing to the overall complexity and level of detail in an image. Such a stepwise visualization conduces to the understanding of the algorithm and validates its effectiveness qualitatively.

Our approach does not incorporate iterative event updates into image deblurring, since the events recovered during iterations are not representative of a complete dynamic scene and therefore do not contain sufficient motion features. Nevertheless, we can still explore whether event denoising leads to better deblurred images by using $\dot{\mathbf{E}}$, which is the output of the last operation, as the prior source. \cref{fig:second_proc} compares the corresponding results. The first image is coarser and involves slight artifacts (\eg, gray patches), and the second is smoother due to lower exposure to event noise. The two settings have little effect on the estimated blur kernel. It shows that denoised events can moderately improve deblurred images.

\input{./src/exdb_limitations}

\section{Limitation Discussions}\label{app:limit}
Although our solution achieves outstanding results in evaluations, its limitation should also be of concern. Since the proposed deblurring mechanism hinges on event priors, it cannot recover the details that are missing from both the blurry image and events, as the sample~\cite{mueggler2017event} given in~\cref{fig:exdb_limitations}. A synthetic kernel blurs the ground truth to make the thin lines around the black shapes faded. Even if the blur kernel can be faithfully estimated, our reconstruction still loses the thin lines since there are almost no events triggered by them. Moreover, as our work focuses on a low-level image enhancement task where monochrome events are sufficient to provide the required features, the algorithm has not been further experimented on color information of events (\eg, the CED dataset~\cite{scheerlinck2019ced}) that is more conducive to high-level vision analysis such as recognition and segmentation.

\end{appendices}

\bibliographystyle{IEEEtran}
\bibliography{refs}
\newpage
\section{Biography Section}
\vskip -25pt plus -1fil
\begin{IEEEbiography}[{\includegraphics[width=1in,height=1.25in,clip,keepaspectratio]{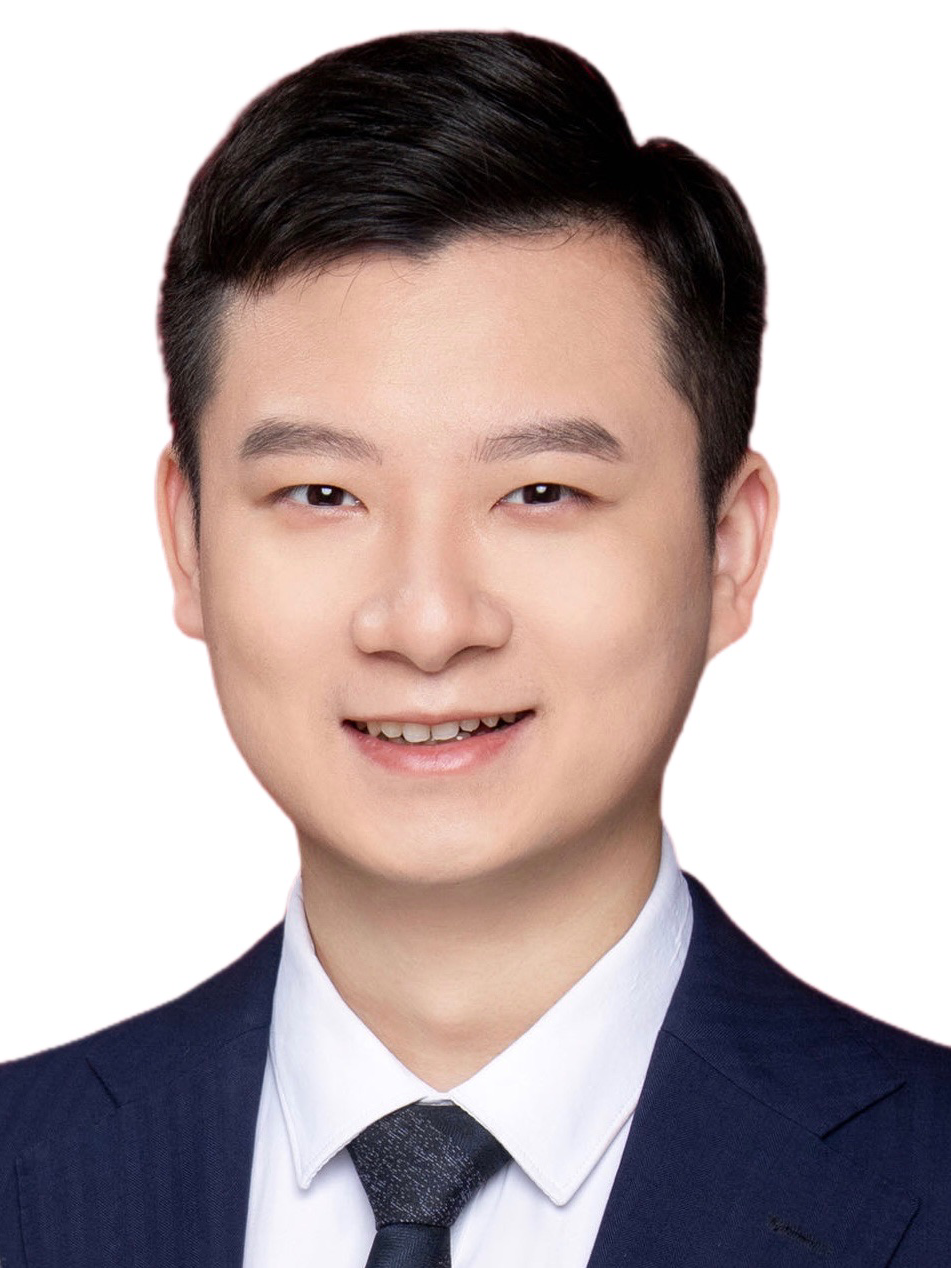}}]{PEI ZHANG} (Graduate Student Member, IEEE) received the B.Eng. degree from Beijing University of Posts and Telecommunications, in 2019, the B.S. degree from Queen Mary University of London, in 2019, and the M.S. degree from University College London, in 2020. He is currently pursuing the Ph.D. degree with the Department of Electrical and Electronic Engineering, The University of Hong Kong. His research interests include computational imaging, neuromorphic imaging and event-based vision.
\end{IEEEbiography}
\vskip -25pt plus -1fil
\begin{IEEEbiography}[{\includegraphics[width=1in,height=1.25in,clip,keepaspectratio]{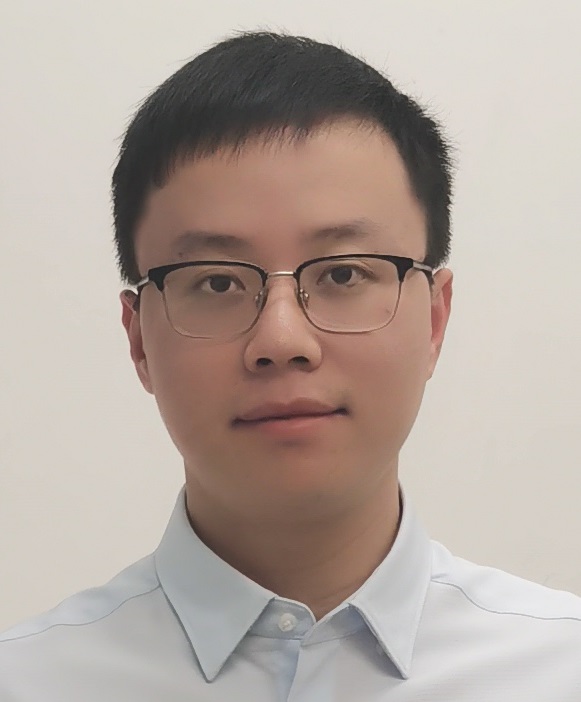}}]{HAOSEN LIU} received the B.S degree and the M.S. degree from the School of Artificial Intelligence and Automation, Huazhong University of Science and Technology, Wuhan, China, in 2016 and 2019, respectively. He is currently pursuing the Ph.D. degree with the Department of Electrical and Electronic Engineering, the University of Hong Kong, Pokfulam, Hong Kong. His research interests include image processing and neuromorphic imaging.
\end{IEEEbiography}
\vskip -25pt plus -1fil
\begin{IEEEbiography}[{\includegraphics[width=1in,height=1.25in,clip,keepaspectratio]{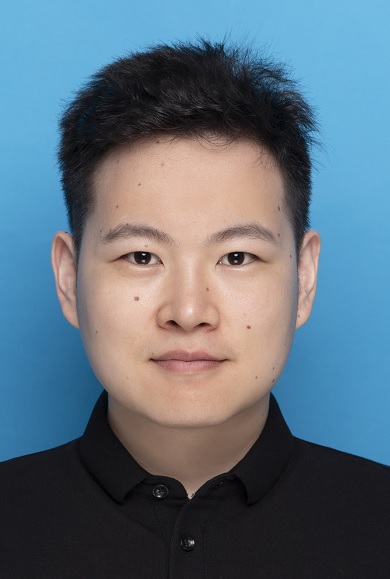}}]{ZHOU GE} received the B.S. degree from Fudan University in 2014, the M.S. degree from Imperial College London in 2015, and the Ph.D. degree from the University of Hong Kong in 2022. He is now an assistant professor at the School of Mechatronic Engineering and Automation, Shanghai University. His research interests include neuromorphic imaging, optical metrology, and digital holography.
\end{IEEEbiography}
\vskip -25pt plus -1fil
\begin{IEEEbiography}[{\includegraphics[width=1in,height=1.25in,clip,keepaspectratio]{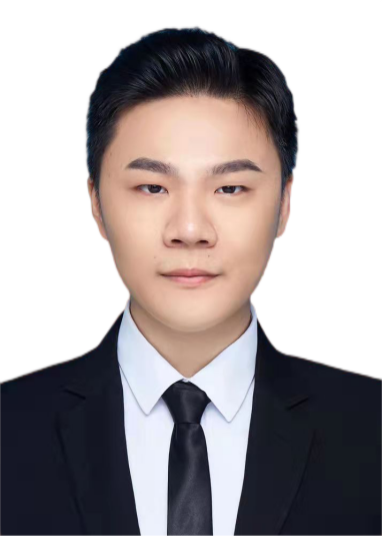}}]{CHUTIAN WANG} received the B.S. degree in Huang Kun Elite Class from the University of Science \& Technology Beijing in 2020, and the M.S. degree in the major of Optics and Photonics in Imperial College London in 2021. He was a research assistant in Zhejiang University until 2022. He is currently working towards his PhD degree with the Department of Electrical and Electronic Engineering, University of Hong Kong. His research interests include computational imaging and neuromorphic imaging.
\end{IEEEbiography}
\vskip -25pt plus -1fil
\begin{IEEEbiography}[{\includegraphics[width=1in,height=1.25in,clip,keepaspectratio]{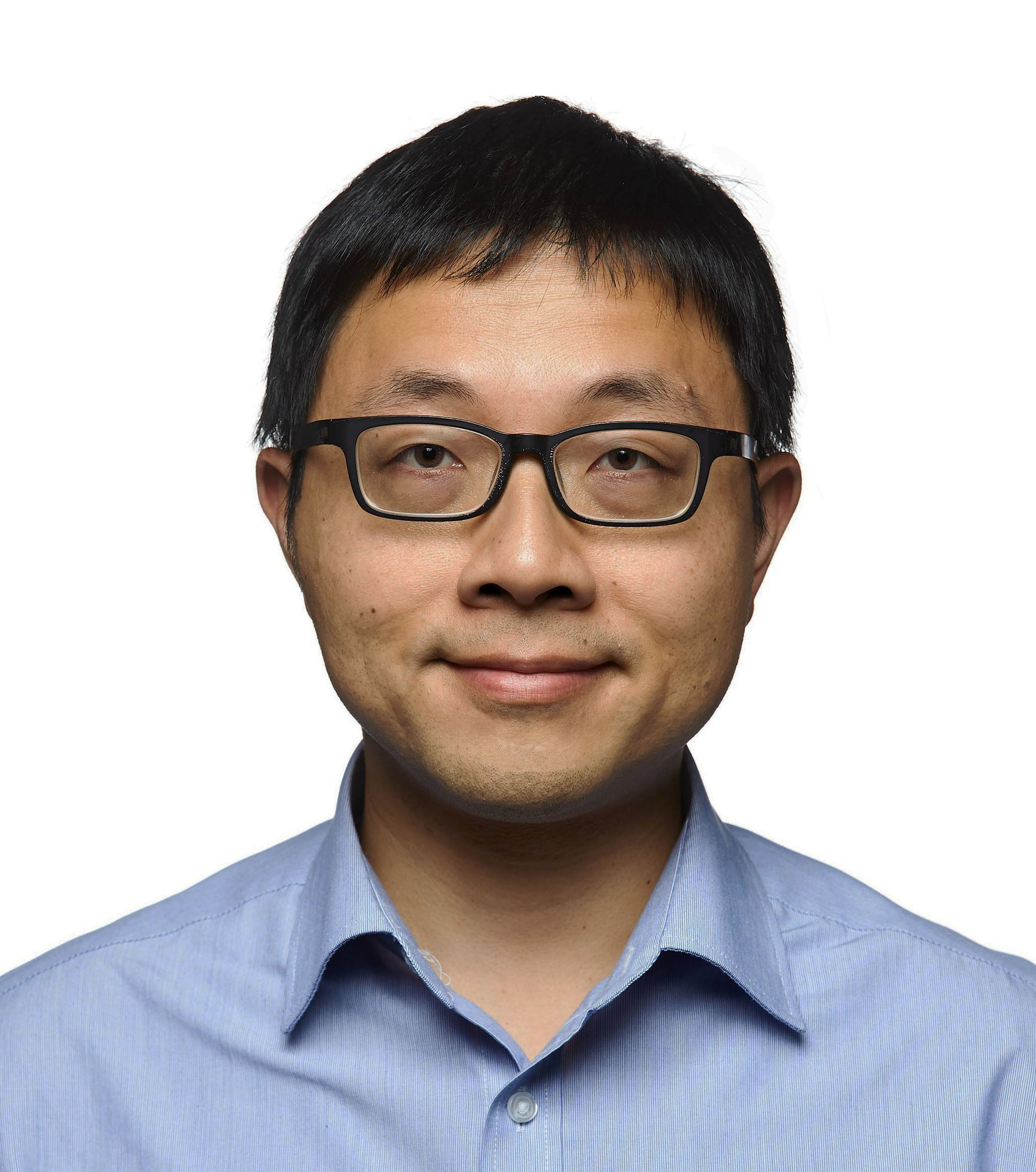}}]{EDMUND Y. LAM} (Fellow, IEEE) received the B.S., M.S., and Ph.D. degrees in electrical engineering from Stanford University. He was a Visiting Associate Professor with the Department of Electrical Engineering and Computer Science, Massachusetts Institute of Technology. He is currently a Professor of electrical and electronic engineering at The University of Hong Kong. He also serves as the Computer Engineering Program Director and a Research Program Coordinator with the AI Chip Center for Emerging Smart Systems. His research interest includes computational imaging algorithms, systems, and applications. He is a fellow of Optica, SPIE, IS\&T, and HKIE, and a Founding Member of the Hong Kong Young Academy of Sciences.
\end{IEEEbiography}

\end{document}

%% file: src/demo_intro.tex
\begin{figure}[t]
    \centering
    \subfloat[The APS with a low frame rate captures a blurry image for the fast-moving badminton (left), and the DVS outputs a lot of noise in the static background (right).]{\frame{\includegraphics[width=\demointro]{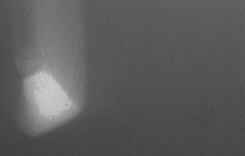}}\hspace{5pt}\stackinset{r}{0pt}{t}{0pt}{\frame{\includegraphics[width=60pt]{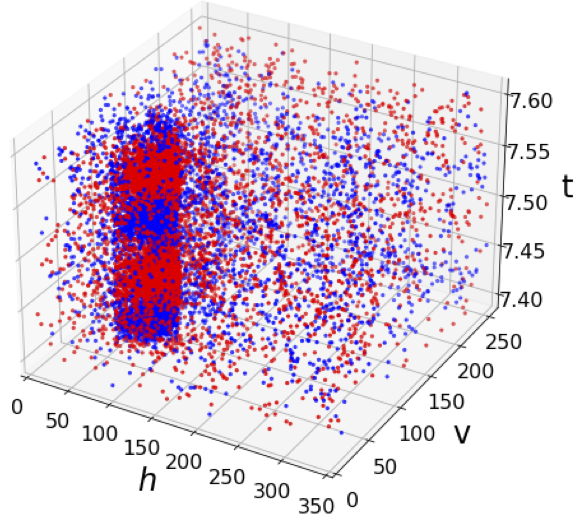}}}{\frame{\includegraphics[width=\demointro]{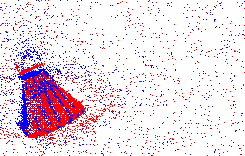}}}}

    \subfloat[The two reconstructions from our approach. The events with high temporal precision assist image deblurring (left), and the estimated sharp image with clear edges steers event denoising (right).]{\frame{\includegraphics[width=\demointro]{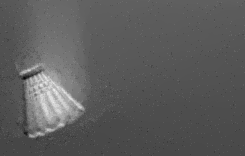}}\hspace{5pt}\stackinset{r}{0pt}{t}{0pt}{\frame{\includegraphics[width=60pt]{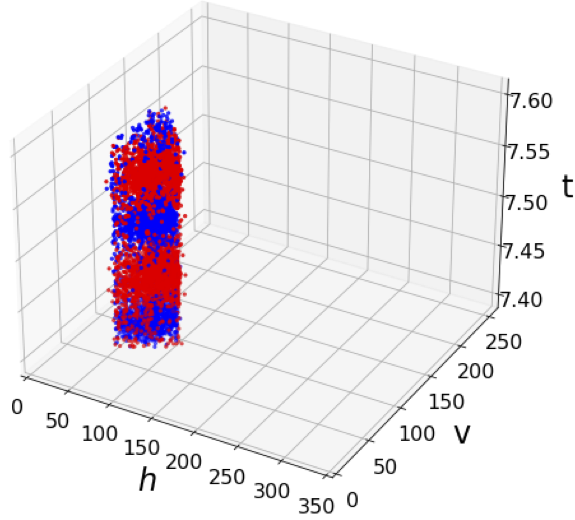}}}{\frame{\includegraphics[width=\demointro]{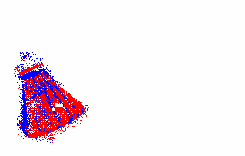}}}}
    \caption{A badminton in free fall. Our algorithm processes a blurry image and a stream of noisy (raw) events, and outputs the corresponding sharp image and denoised events. Events are shown in two-dimensional (2D) and 3D views.}
    \label{fig:demo_intro}
\end{figure}

%% file: src/demo.tex
\begin{figure}[t]
    \centering
    \subfloat[Image $\mathbf{S}$~\cite{mueggler2017event}]{\frame{\includegraphics[width=\demo]{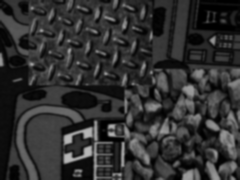}}}
    \hspace{1pt}
    \subfloat[Gradient $\nabla_{\mathbf{x}} \mathbf{S}$]{\frame{\includegraphics[width=\demo]{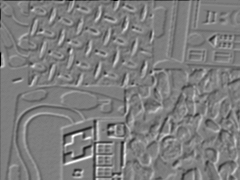}}}
    \hspace{1pt}
    \subfloat[Event Prior $\mathbf{I}_\tau(t)$]{\frame{\includegraphics[width=\demo]{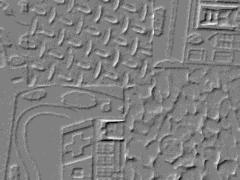}}}
    \caption{Both gradients and events represent high-frequency features of a scene, and the latter can be the ground truth of a latent sharp image in the gradient domain.}
    \label{fig:demo}
\end{figure}

%% file: algorithms/alg.tex
\begin{algorithm*}[t]
    \label{alg}
    \DontPrintSemicolon
    \SetKwInOut{Input}{Input}
    \SetKwInOut{Output}{Output}
    \SetKw{AND}{\textbf{and}}
    \caption{Joint Reconstruction of the Blur-free Image and Noise-robust Events}
    
    \Input{$\mathbf{B}$, $\mathbf{E}$}
    \Output{$\mathbf{S}, \dot{\mathbf{E}}$}
    Initialize $l = 0$, and $\mathbf{k}^{(l)}$ from the coarser level.\tcp*{$l$ indexes iterations}
    Initialize $\mathbf{I}_\tau(t)$ from $\mathbf{E}$ by~\Cref{eq:event_,eq:intensity}.\;
    Let $\mathbf{S}^{(l-1)} = \mathbf{B}$, and $\dot{\mathbf{E}}^{(l)} = \varnothing$.\;
    
    \While(\tcp*[f]{$l_{max}$ defines the maximum iteration}){$l \leqslant l_{max}$}
    {
        $\gamma = 2\beta$. \;
        \While(\tcp*[f]{$\gamma_{max}$ limits the update of $\mathbf{S}$}){$\gamma \leqslant \gamma_{max}$}
        {
            With $\gamma, \beta$, update $\mathbf{z}^{(l)}$ by~\Cref{eq:sol_z}.\;
            With $\mathbf{k}^{(l)}, \mathbf{z}^{(l)}$, $\mathbf{I}_\tau(t)$, update $\mathbf{S}^{(l)}$ by~\Cref{eq:sol_s}.\;
            $\gamma = 2\gamma$.\;
        }
        With $\mathbf{S}^{(l)}$, update $\mathbf{g}^{(l)}$ by \Cref{eq:sol_g}.\;
        $\tilde{\mathbf{E}}(t) = \frac{\partial}{\partial t}\mathbf{I}_\tau(t)$.\;
        With $\tilde{\mathbf{E}}(t)$ and $\mathbf{g}^{(l)}$, update $\dot{\mathbf{E}}^{(l)}$ by~\Cref{eq:denoised_e}.\;
        With $\dot{\mathbf{E}}^{(l)}$, update $\mathbf{G}^{(i)}_{\mu, \nu}$ by~\Cref{eq:neighbors}.\tcp*{can be any search method}
        $\dot{\mathbf{E}}^{(l)} = \dot{\mathbf{E}}^{(l)} \cup \mathbf{G}^{(i)}_{\mu, \nu}$.\;
        With $\mathbf{S}^{(l)}$, update $\mathbf{k}^{(l+1)}$ by~\Cref{eq:estimated_k}.\;
        $l = l + 1$.\;
    }
\end{algorithm*}

%% file: src/exdb_uniform.tex
\begin{figure*}[t]
    \centering
    \subfloat[Pseudo Ground Truth]{\frame{\includegraphics[width=\exdbuniform]{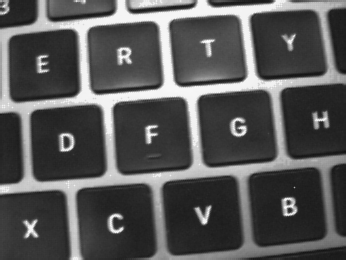}}}
    \hspace{6pt}%
    \subfloat[Blurry Image]{\stackinset{r}{-1pt}{b}{0pt}{\includegraphics[width=60pt]{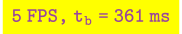}}{\frame{\includegraphics[width=\exdbuniform]{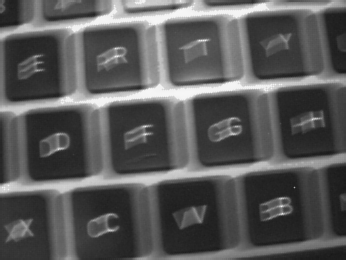}}}}
    \hspace{6pt}%
    \subfloat[Event Prior $\mathbf{I}_\tau(t)$ with $\tau = 6$ \si{\ms}]{\frame{\includegraphics[width=\exdbuniform]{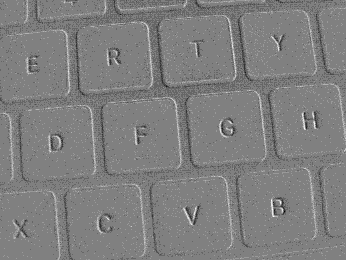}}}
    \hspace{6pt}%
    \subfloat[Ours]{\frame{\includegraphics[width=\exdbuniform]{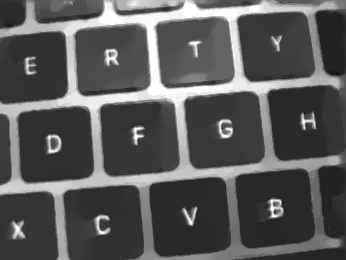}}}
    \hspace{6pt}%
    \vspace{-6pt}
    \\
    \subfloat[Bai~\etal~\cite{bai2018graph}]{\frame{\includegraphics[width=\exdbuniform]{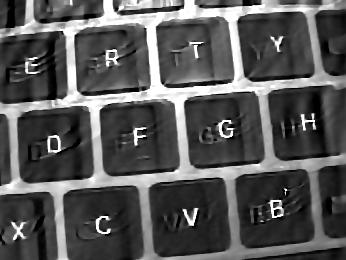}}}
    \hspace{6pt}%
    \subfloat[Scheerlinck~\etal~\cite{scheerlinck2018continuous}]{\frame{\includegraphics[width=\exdbuniform]{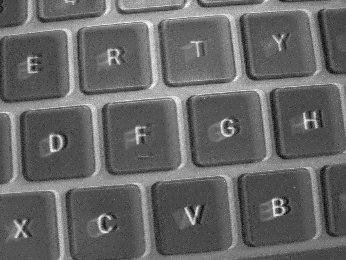}}}
    \hspace{6pt}%
    \subfloat[Pan~\etal~\cite{pan2019bringing,pan2022high}]{\frame{\includegraphics[width=\exdbuniform]{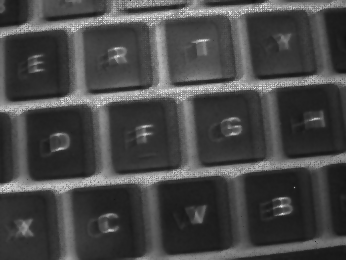}}}
    \hspace{6pt}%
    \subfloat[Rebecq~\etal~\cite{rebecq2019events,rebecq2021high}]{\frame{\includegraphics[width=\exdbuniform]{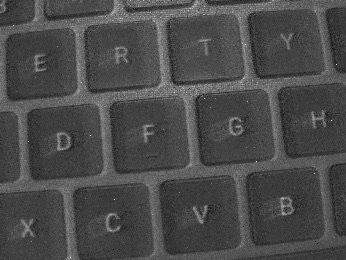}}}
    \hspace{6pt}%
    \caption{Visual comparisons on deblurring an image with uniform motion blur. (a) A sharp image at a certain timestamp serves as a ground truth for reference. (b) The blurry image to be processed. (c) The used event prior. (d) Our reconstruction has more faithful shapes and shades of gray. (e) The image-based method has distortions and ringing artifacts in some pixels. (f)--(h) The results of the competing event-based approaches, where (f) leaves blur residues and (g), (h) involve a mass of gray flecks.}
    \label{fig:exdb_uniform}
\end{figure*}

%% file: src/uniform.tex
\begin{figure*}[t]
    \centering
    \subfloat[Ground Truth]{\frame{\includegraphics[width=\exdbuniform]{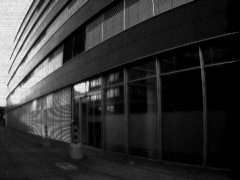}}}
    \hspace{6pt}%
    \subfloat[Blurry Image]{\stackinset{l}{0pt}{b}{0pt}{\includegraphics[width=32pt]{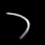}}{\stackinset{l}{0pt}{t}{0pt}{\frame{\includegraphics[width=60pt]{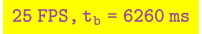}}}
    {\frame{\includegraphics[width=\exdbuniform]{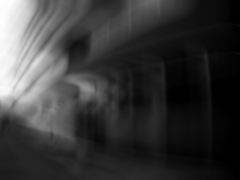}}}}}
    \hspace{6pt}%
    \subfloat[Event Prior $\mathbf{I}_\tau(t)$ with $\tau = 10$ \si{\ms}]{\frame{\includegraphics[width=\exdbuniform]{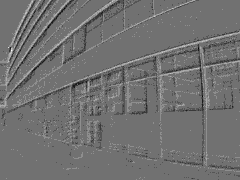}}}
    \hspace{6pt}%
    \subfloat[Ours]{\stackinset{l}{0pt}{b}{0pt}{\frame{\includegraphics[width=32pt]{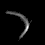}}}{\frame{\includegraphics[width=\exdbuniform]{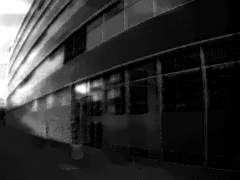}}}}
    \hspace{6pt}%
    \vspace{-6pt}
    \\
    \subfloat[Pan~\etal~\cite{pan2019phase}]{\frame{\includegraphics[width=\exdbuniform]{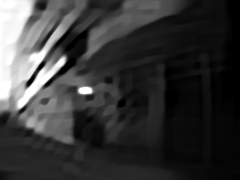}}}
    \hspace{6pt}%
    \subfloat[Scheerlinck~\etal~\cite{scheerlinck2018continuous}]{\frame{\includegraphics[width=\exdbuniform]{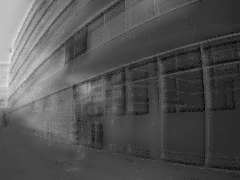}}}
    \hspace{6pt}%
    \subfloat[Pan~\etal~\cite{pan2019bringing,pan2022high}]{\frame{\includegraphics[width=\exdbuniform]{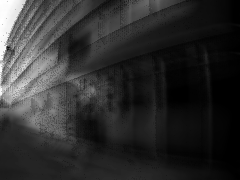}}}
    \hspace{6pt}%
    \subfloat[Sun~\etal~\cite{sun2022event}]{\frame{\includegraphics[width=\exdbuniform]{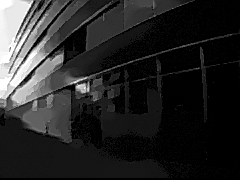}}}
    \hspace{6pt}%
    \caption{Visual comparisons on deblurring an image with synthetic uniform motion blur. (a) The raw sharp image. (b) The blurry image is the result of a convolution of (a) with the given blur kernel. (c) The used event prior. (d) Our approach reverses the blurring process accurately and enhances the image clarity. (e) The image-based method cannot perform a faithful restoration. (f)--(h) The results of the event-based competitors, where (g) is still blurry, and (h) with a learning-based fashion recovers the overall structure but loses some details.}
    \label{fig:uniform}
\end{figure*}

%% file: src/exdb_nonuniform.tex
\begin{figure*}[t]
    \centering
    \subfloat[Pseudo Ground Truth]{\frame{\includegraphics[width=\exdbnonuniform]{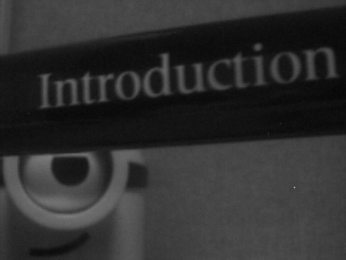}}}
    \hspace{6pt}%
    \subfloat[Blurry Image]{\stackinset{r}{-1pt}{b}{0pt}{\includegraphics[width=60pt]{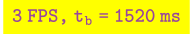}}{\frame{\includegraphics[width=\exdbnonuniform]{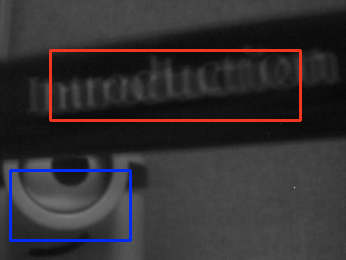}}}}
    \hspace{6pt}%
    \subfloat[Event Prior $\mathbf{I}_\tau(t)$ with $\tau = 6$ \si{\ms}]{\frame{\includegraphics[width=\exdbnonuniform]{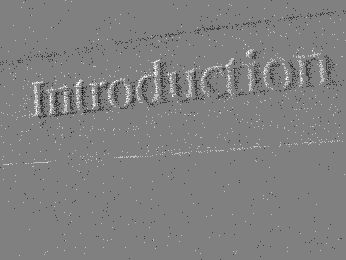}}}
    \hspace{6pt}%
    \subfloat[Ours]{\frame{\includegraphics[width=\exdbnonuniform]{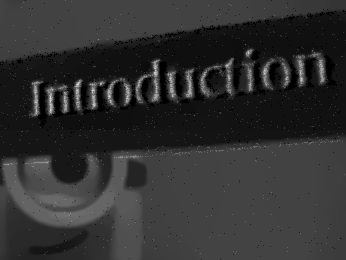}}}
    \hspace{6pt}%
    \\\vspace{-6pt}
    \subfloat[Dong~\etal~\cite{dong2017blind}]{\frame{\includegraphics[width=\exdbnonuniform]{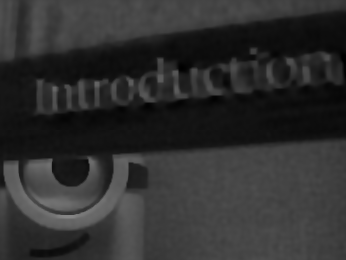}}}
    \hspace{6pt}%
    \subfloat[Scheerlinck~\etal~\cite{scheerlinck2018continuous}]{\frame{\includegraphics[width=\exdbnonuniform]{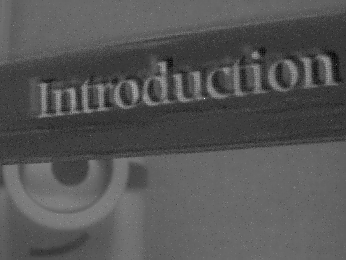}}}
    \hspace{6pt}%
    \subfloat[Pan~\etal~\cite{pan2019bringing,pan2022high}]{\frame{\includegraphics[width=\exdbnonuniform]{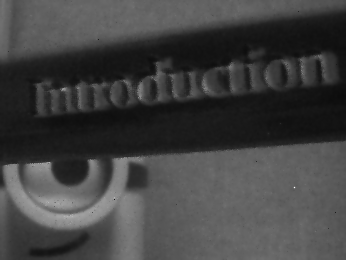}}}
    \hspace{6pt}%
    \subfloat[Rebecq~\etal~\cite{rebecq2019events,rebecq2021high}]{\frame{\includegraphics[width=\exdbnonuniform]{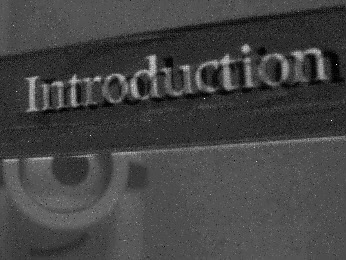}}}
    \hspace{6pt}%
    \caption{Visual comparisons on deblurring an image with non-uniform motion blur. (a) A sharp image at a certain timestamp serves as a ground truth for reference. (b) The raw image, where the blurry foreground to be processed is highlighted with a red box, and the static background is marked by a blue box. (c) The used event prior. (d) Our image has clearer visualization of the deblurred texts. (e) The image-based algorithm still leaves severe blur residues. (f)--(h) The reconstructions of the competing event-based methods present different degrees of artifacts.}
    \label{fig:exdb_nonuniform}
\end{figure*}

%% file: src/nonuniform.tex
\begin{figure*}[t]
    \centering
    \subfloat[Pseudo Ground Truth]{\frame{\includegraphics[width=\exdbnonuniform, height=\nonuniform]{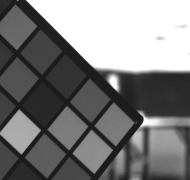}}}
    \hspace{6pt}%
    \subfloat[Blurry Image]{\stackinset{l}{0pt}{b}{0pt}{\includegraphics[width=25pt]{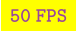}}{\frame{\includegraphics[width=\exdbnonuniform, height=\nonuniform]{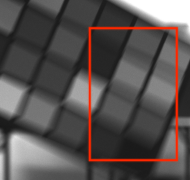}}}}
    \hspace{6pt}%
    \subfloat[Event Prior $\mathbf{I}_\tau(t)$ with $\tau = 3$ \si{\ms}]{\frame{\includegraphics[width=\exdbnonuniform, height=\nonuniform]{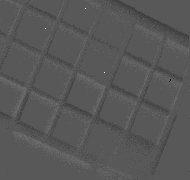}}}
    \hspace{6pt}%
    \subfloat[Ours]{\frame{\includegraphics[width=\exdbnonuniform, height=\nonuniform]{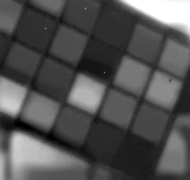}}}
    \hspace{6pt}%
    \\\vspace{-6pt}
    \subfloat[Bai~\etal~\cite{bai2018graph}]{\frame{\includegraphics[width=\exdbnonuniform, height=\nonuniform]{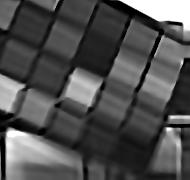}}}
    \hspace{6pt}%
    \subfloat[Scheerlinck~\etal~\cite{scheerlinck2018continuous}]{\frame{\includegraphics[width=\exdbnonuniform, height=\nonuniform]{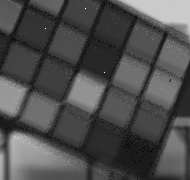}}}
    \hspace{6pt}%
    \subfloat[Pan~\etal~\cite{pan2019bringing,pan2022high}]{\frame{\includegraphics[width=\exdbnonuniform, height=\nonuniform]{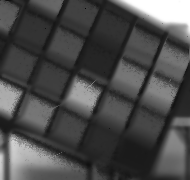}}}
    \hspace{6pt}%
    \subfloat[Sun~\etal~\cite{sun2022event}]{\frame{\includegraphics[width=\exdbnonuniform, height=\nonuniform]{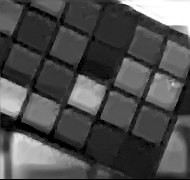}}}
    \hspace{6pt}%
    \caption{Visual comparisons on deblurring an image with non-uniform motion blur. (a) A sharp image at a certain timestamp serves as a ground truth for reference. (b) The blurry image, where the area of concern has blocks with different contrast. (c) The used event prior. (d) Our approach faithfully recovers the lines, edges and shade of gray. (e) Obvious blur residues appear in the result of the image-based method. (f)--(h) The images of the event-based competitors, where (h) exhibits ringing artifacts.}
    \label{fig:nonuniform}
\end{figure*}

%% file: src/exdb_extreme.tex
\begin{figure*}[t]
    \centering
    \hspace{-9.5pt}%
    \subfloat[Raw Images]{
    \begin{tabular}[b]{c}
        \stackinset{l}{0pt}{t}{0pt}{\includegraphics[width=60pt]{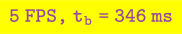}}{\frame{\includegraphics[width=\exdbextreme]{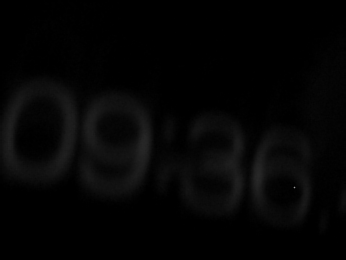}}}\\
        \stackinset{l}{0pt}{t}{0pt}{\includegraphics[width=60pt]{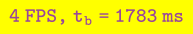}}{\frame{\includegraphics[width=\exdbextreme]{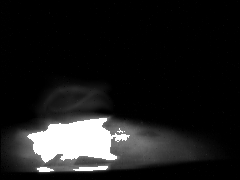}}}\\
        \stackinset{l}{0pt}{t}{0pt}{\includegraphics[width=60pt]{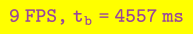}}{\stackinset{r}{0pt}{b}{0pt}{\frame{\includegraphics[width=50pt]{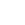}}}
    {\frame{\includegraphics[width=\exdbextreme]{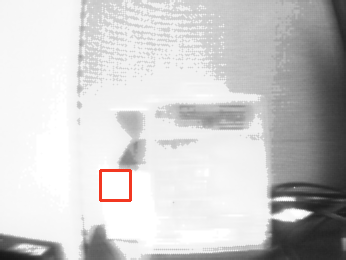}}}}\\
        \stackinset{l}{0pt}{t}{0pt}{\includegraphics[width=60pt]{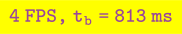}}{\stackinset{r}{0pt}{b}{0pt}{\frame{\includegraphics[width=50pt]{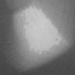}}}
    {\frame{\includegraphics[width=\exdbextreme]{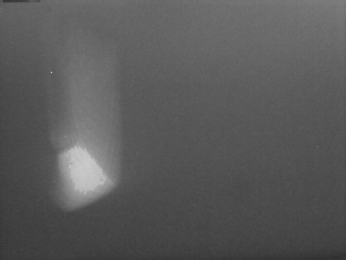}}}}
    \end{tabular}}\hspace{-9.5pt}%
    \subfloat[Ours]{
    \begin{tabular}[b]{c}%
        \frame{\includegraphics[width=\exdbextreme]{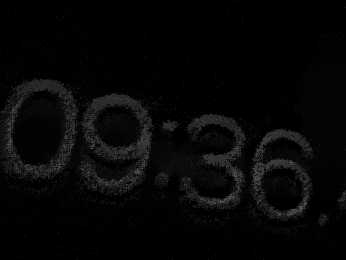}}\\
        \frame{\includegraphics[width=\exdbextreme]{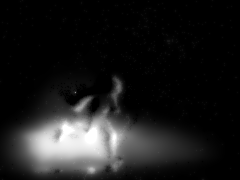}}\\
        \stackinset{r}{0pt}{b}{0pt}{\frame{\includegraphics[width=50pt]{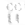}}}{\frame{\includegraphics[width=\exdbextreme]{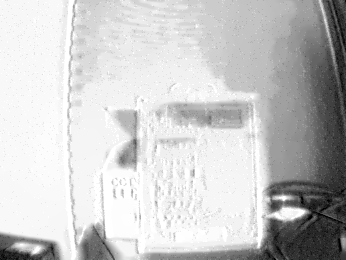}}}\\
        \stackinset{r}{0pt}{b}{0pt}{\frame{\includegraphics[width=50pt]{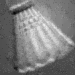}}}{\frame{\includegraphics[width=\exdbextreme]{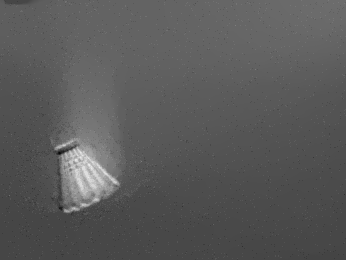}}}
    \end{tabular}}\hspace{-9.5pt}%
    \subfloat[Scheerlinck~\etal~\cite{scheerlinck2018continuous}]{
    \begin{tabular}[b]{c}
        \frame{\includegraphics[width=\exdbextreme]{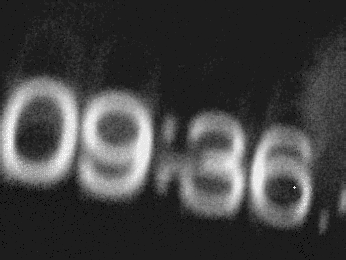}}\\
        \frame{\includegraphics[width=\exdbextreme]{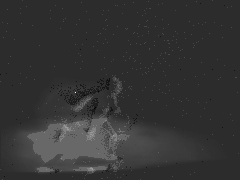}}\\
        \stackinset{r}{0pt}{b}{0pt}{\frame{\includegraphics[width=50pt]{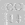}}}{\frame{\includegraphics[width=\exdbextreme]{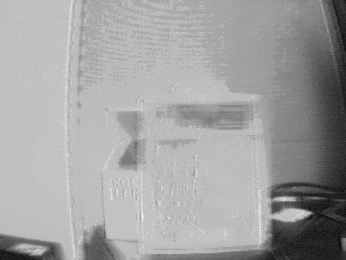}}}\\
        \stackinset{r}{0pt}{b}{0pt}{\frame{\includegraphics[width=50pt]{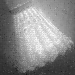}}}{\frame{\includegraphics[width=\exdbextreme]{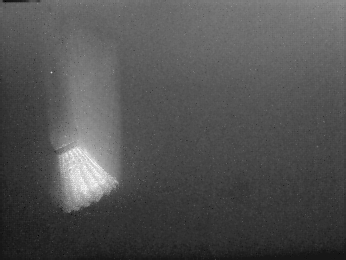}}}
    \end{tabular}}\hspace{-9.5pt}%
    \subfloat[Pan~\etal~\cite{pan2019bringing,pan2022high}]{
    \begin{tabular}[b]{c}%
        \frame{\includegraphics[width=\exdbextreme]{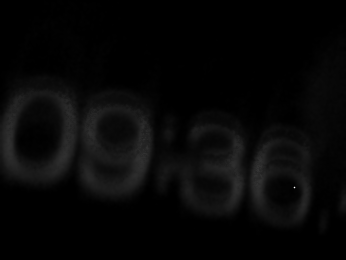}}\\
        \frame{\includegraphics[width=\exdbextreme]{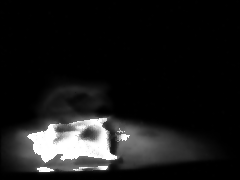}}\\
        \stackinset{r}{0pt}{b}{0pt}{\frame{\includegraphics[width=50pt]{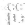}}}{\frame{\includegraphics[width=\exdbextreme]{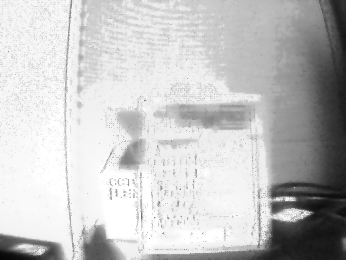}}}\\
        \stackinset{r}{0pt}{b}{0pt}{\frame{\includegraphics[width=50pt]{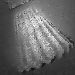}}}{\frame{\includegraphics[width=\exdbextreme]{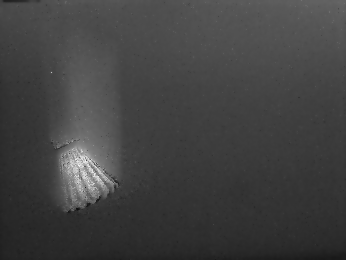}}}
    \end{tabular}}
    \caption{Reconstructing the images captured in challenging scenes. (a) The raw images suffer from information loss due to the nature of the APS. (b) Our results, which are based on around $14\,000$ (collected in $2$ \si{\ms}), $3\,000$ ($4$ \si{\ms}), $28\,000$ ($10$ \si{\ms}) and $11\,000$ ($20$ \si{\ms}) events respectively, have clearer shapes and more delicate textures. (c)--(d) The results of the state-of-the-art event-based methods, whose shades of gray and textures are less faithful to the original.}
    \label{fig:exdb_extreme}
\end{figure*}

%% file: src/exdb_gt_tab.tex
\begin{table}[t]
\caption{Quantitative comparisons on blind image deblurring.}
\label{tab:exdb_gt_tab}
\centering
\begin{tabular}{llccc}
    \toprule
    \textbf{Type} &\textbf{Algorithm} & \multicolumn{3}{c}{\textbf{Evaluation}} \\ \midrule
    & &\textbf{MSE} $\downarrow$& \textbf{LPIPS} $\downarrow$ & \textbf{SSIM} $\uparrow$ \\ \cmidrule(lr){3-5}
    &Synthetic Blur&  5.34 & 22.52 & 0.17\\\midrule
    \multirow{4}{*}{\texttt{Image}}&Dong~\etal~\cite{dong2017blind} & 3.59 & 12.62 & 0.28 \\
    &Bai~\etal~\cite{bai2018graph} & 4.05 &  13.05 & 0.25\\
    &Chen~\etal~\cite{chen2019blind} & 3.34 & 15.22 & 0.35\\
    &Pan~\etal~\cite{pan2019phase} & 3.67 & 12.87 & 0.32 \\\midrule
    \multirow{1}{*}{\texttt{Event}}&Rebecq~\etal~\cite{rebecq2019events,rebecq2021high} & 2.98 & 6.53 & 0.41\\\midrule
    \multirow{4}{*}{\texttt{Fusion}}&Scheerlinck~\etal~\cite{scheerlinck2018continuous} & 3.04 & 10.14 & 0.39\\
    &Pan~\etal~\cite{pan2019bringing,pan2022high} & 2.87 & 9.26 & 0.35 \\ 
    &Wang~\etal~\cite{wang2021asynchronous} & 3.22 & 7.79 & 0.39\\
    &Sun~\etal~\cite{sun2022event} & 2.11 & 5.83 & 0.45\\ \midrule
    &\textbf{Ours} & \textbf{1.95} & \textbf{5.72} & \textbf{0.47}\\
    \bottomrule
\end{tabular}
\end{table}

%% file: src/exdn_sample.tex
\begin{figure*}[t]
    \centering
    \hspace{-9.5pt}%
    \subfloat[Raw Events]{
    \begin{tabular}[b]{c}
        \stackinset{r}{-1pt}{t}{0pt}{\includegraphics[width=80pt]{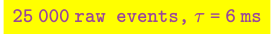}}{\stackinset{r}{0pt}{b}{0pt}{\frame{\includegraphics[width=45pt]{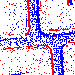}}}{\frame{\includegraphics[width=\exdnsample]{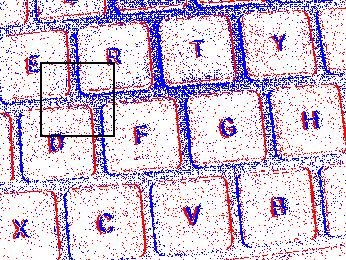}}}}\\
        \stackinset{r}{-1pt}{b}{0pt}{\includegraphics[width=80pt]{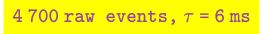}}{\stackinset{l}{0pt}{b}{0pt}{\frame{\includegraphics[height=48pt]{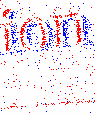}}}{\frame{\includegraphics[width=\exdnsample]{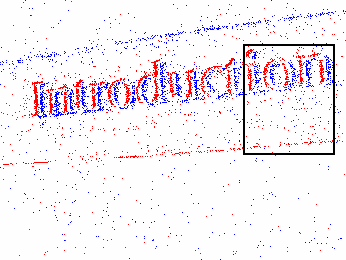}}}}\\
        \stackinset{r}{-1pt}{t}{0pt}{\includegraphics[width=80pt]{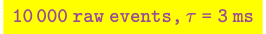}}{\stackinset{r}{0pt}{b}{0pt}{\frame{\includegraphics[height=40pt]{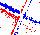}}}{\frame{\includegraphics[width=\exdnsample,height=93pt]{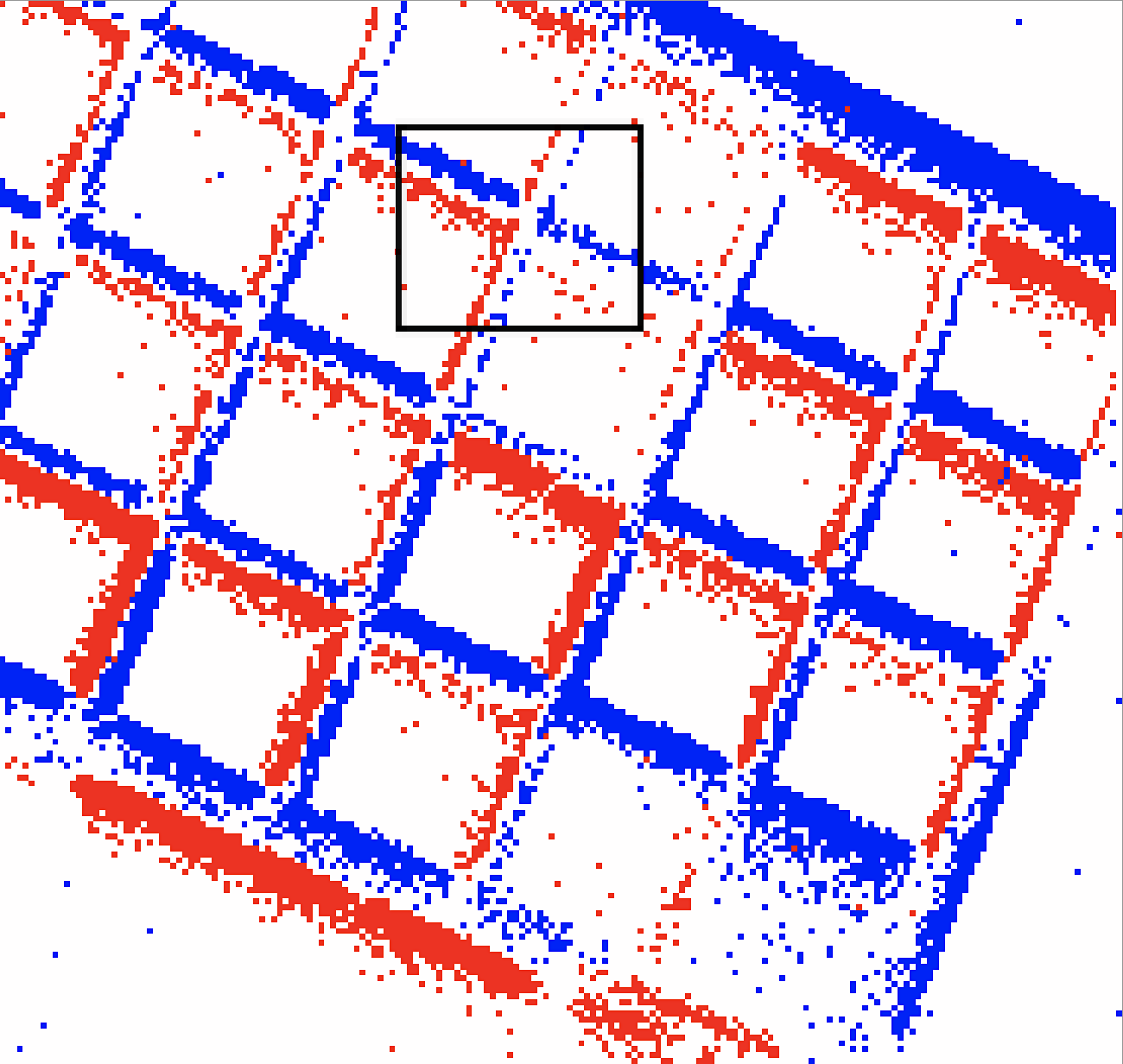}}}}
    \end{tabular}}\hspace{-9.5pt}%
    \subfloat[Ours]{
    \begin{tabular}[b]{c}%
        \stackinset{r}{0pt}{b}{0pt}{\frame{\includegraphics[width=45pt]{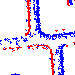}}}{\frame{\includegraphics[width=\exdnsample]{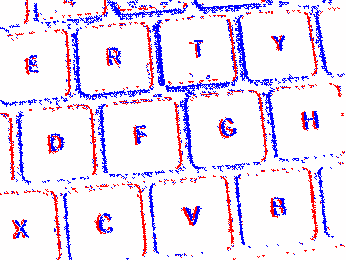}}}\\
        \stackinset{l}{0pt}{b}{0pt}{\frame{\includegraphics[height=48pt]{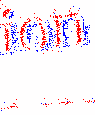}}}{\frame{\includegraphics[width=\exdnsample]{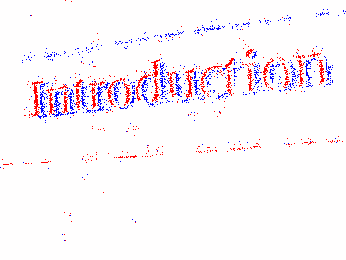}}}\\
        \stackinset{r}{0pt}{b}{0pt}{\frame{\includegraphics[height=40pt]{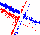}}}{\frame{\includegraphics[width=\exdnsample,height=93pt]{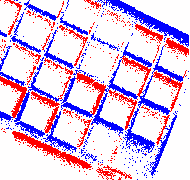}}}\\
    \end{tabular}}\hspace{-9.5pt}%
    \subfloat[Feng~\etal~\cite{feng2020event}]{
    \begin{tabular}[b]{c}%
        \stackinset{r}{0pt}{b}{0pt}{\frame{\includegraphics[width=45pt]{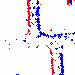}}}{\frame{\includegraphics[width=\exdnsample]{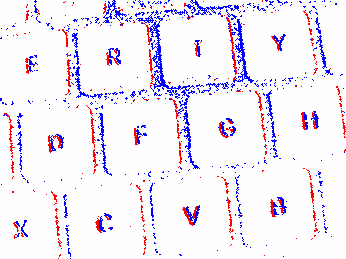}}}\\
        \stackinset{l}{0pt}{b}{0pt}{\frame{\includegraphics[height=48pt]{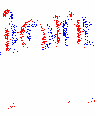}}}{\frame{\includegraphics[width=\exdnsample]{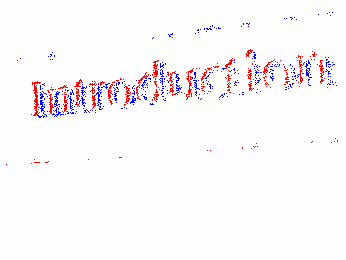}}}\\
        \stackinset{r}{0pt}{b}{0pt}{\frame{\includegraphics[height=40pt]{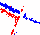}}}{\frame{\includegraphics[width=\exdnsample,height=93pt]{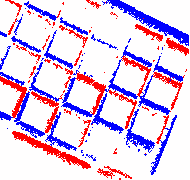}}}
    \end{tabular}}\hspace{-9.5pt}%
    \subfloat[Wu~\etal~\cite{wu2020probabilistic}]{
    \begin{tabular}[b]{c}%
        \stackinset{r}{0pt}{b}{0pt}{\frame{\includegraphics[width=45pt]{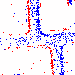}}}{\frame{\includegraphics[width=\exdnsample]{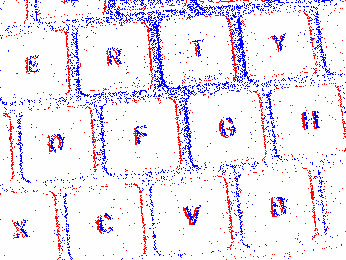}}}\\
        \stackinset{l}{0pt}{b}{0pt}{\frame{\includegraphics[height=48pt]{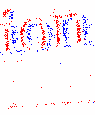}}}{\frame{\includegraphics[width=\exdnsample]{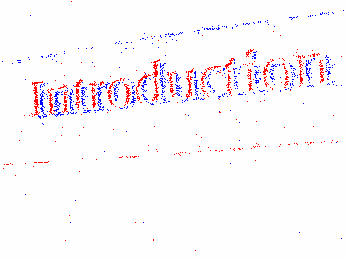}}}\\
        \stackinset{r}{0pt}{b}{0pt}{\frame{\includegraphics[height=40pt]{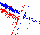}}}{\frame{\includegraphics[width=\exdnsample,height=93pt]{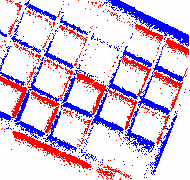}}}
    \end{tabular}}\hspace{-9.5pt}%
    \caption{Noise removal of the raw events triggered by the scenes of~\cref{fig:exdb_uniform,fig:exdb_nonuniform,fig:nonuniform}. We use an event frame to visualize the stream. (a) Raw noisy data, where the marked zones with zoom-in consist of events and noise. (b) Our results suppress most of noise and faithfully retain events. (c)--(d) The results of the counterparts, where (c) loses a significant number of events, and (d) keeps events as well as a small amount of sparse noise.}
    \label{fig:exdn_sample}
\end{figure*}

%% file: src/exdn_gt_tab.tex
\begin{table}[t]
\caption{Quantitative comparisons on neuromorphic noise removal.}
\label{tab:exdn_gt_tab}
\centering
\begin{tabular}{llcccc}
    \toprule
    \textbf{Sample} & \textbf{Algorithm} & \multicolumn{4}{c}{\textbf{Evaluation}} \\ \midrule
     & & \textbf{TPR} & \textbf{FPR} & \textbf{PPV} & \textbf{ACC} $\uparrow$\\ \cmidrule(lr){3-6}
    \multirow{5}{*}{\texttt{Running}} & NN-Filter~\cite{czech2016evaluating} & 0.56 & 0.49 & 0.70 & 0.54\\
    &Feng~\etal~\cite{feng2020event} & 0.59 & 0.19 & 0.86 & 0.66\\
    &Wu~\etal~\cite{wu2020probabilistic} & 0.64 & 0.39 & 0.77 & 0.63\\
    &EDnCNN~\cite{baldwin2020event} & 0.73 & 0.46 & 0.76 & 0.66\\ \cmidrule(lr){2-6}
    &\textbf{Ours} & \textbf{0.76} & \textbf{0.30} & \textbf{0.84} & \textbf{0.74}\\\midrule
    \multirow{5}{*}{\texttt{Shapes}} & NN-Filter~\cite{czech2016evaluating} & 0.84 & 0.29 & 0.85 & 0.80\\
    &Feng~\etal~\cite{feng2020event} & 0.82 & 0.04 & 0.97 & 0.86\\
    &Wu~\etal~\cite{wu2020probabilistic} & 0.85 & 0.26 & 0.87 & 0.81\\
    &EDnCNN~\cite{baldwin2020event} & 0.90 & 0.27 & 0.87 & 0.84\\ \cmidrule(lr){2-6}
    &\textbf{Ours} & \textbf{0.93} & \textbf{0.20} & \textbf{0.90} & \textbf{0.89}\\
    \bottomrule
\end{tabular}
\end{table}

%% file: src/exdn_fast_slow.tex
\begin{figure*}[t]
    \centering
    \hspace{-9.5pt}%
    \subfloat[Raw Events]{
    \begin{tabular}[b]{c}
        \stackinset{l}{-1pt}{t}{0pt}{\includegraphics[width=80pt]{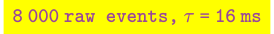}}{\frame{\includegraphics[width=\exdnfastslow]{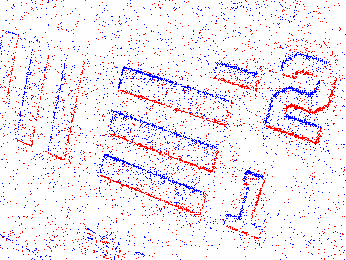}}}\\
        \stackinset{l}{-1pt}{t}{0pt}{\includegraphics[width=80pt]{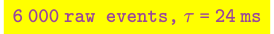}}{\frame{\includegraphics[width=\exdnfastslow]{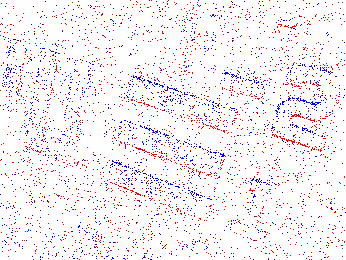}}}
    \end{tabular}}\hspace{-9.5pt}%
    \subfloat[Ours]{
    \begin{tabular}[b]{c}%
        \frame{\includegraphics[width=\exdnfastslow]{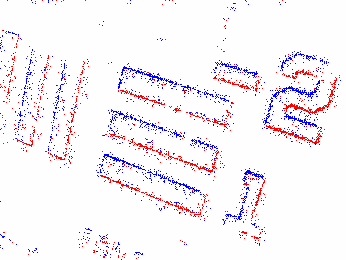}}\\
        \frame{\includegraphics[width=\exdnfastslow]{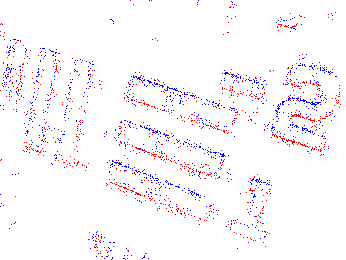}}
    \end{tabular}}\hspace{-9.5pt}%
    \subfloat[Feng~\etal~\cite{feng2020event}]{
    \begin{tabular}[b]{c}%
        \frame{\includegraphics[width=\exdnfastslow]{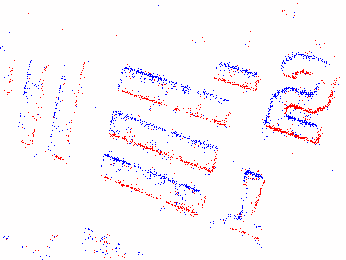}}\\
        \frame{\includegraphics[width=\exdnfastslow]{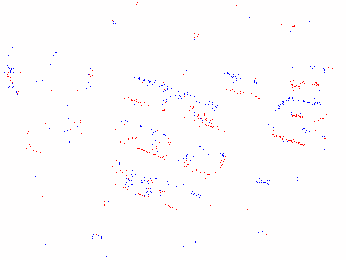}}
    \end{tabular}}\hspace{-9.5pt}%
    \subfloat[Wu~\etal~\cite{wu2020probabilistic}]{
    \begin{tabular}[b]{c}%
        \frame{\includegraphics[width=\exdnfastslow]{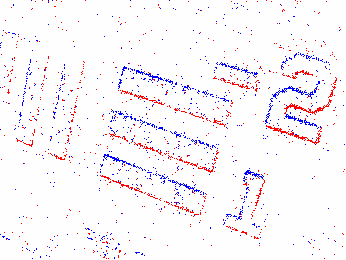}}\\
        \frame{\includegraphics[width=\exdnfastslow]{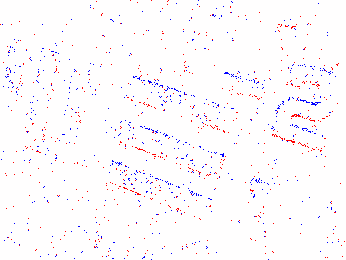}}
    \end{tabular}}\hspace{-9.5pt}%
    \caption{Noise removal of the raw events from a fast-moving object (1st row) and a slow-moving object (2nd row). (a) The events of the fast-moving object are much denser but as sparse as noise when the object barely moves. (b) Our results have considerably reduced noise and faithfully preserved events. (c) For the slow-moving case, the density-based mechanism mistakes events for sparse noise since there is no significant difference in density. (d) The results still have observable noise in the background.}
    \label{fig:exdn_fast_slow}
\end{figure*}

%% file: src/exdb_evsnum.tex
\begin{figure}[t]
    \centering
    \subfloat[The results of Pan~\etal~\cite{pan2019bringing,pan2022high} (left) and ours (right) based on $4\,700$ events.]{\frame{\includegraphics[width=\exdbevsnum]{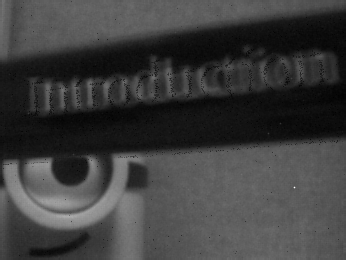}}\hspace{5pt}\frame{\includegraphics[width=\exdbevsnum]{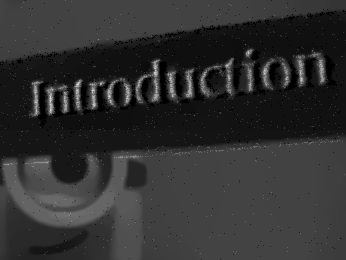}}}
    \\
    \vspace{-6pt}
    \subfloat[The results of Scheerlinck~\etal~\cite{scheerlinck2018continuous} (left) and ours (right) based on $25\,000$ events.]{\frame{\includegraphics[width=\exdbevsnum]{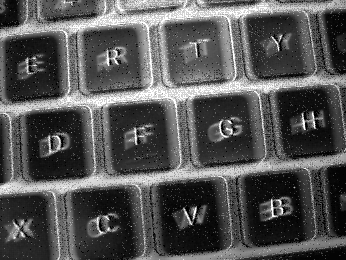}}\hspace{5pt}\frame{\includegraphics[width=\exdbevsnum]{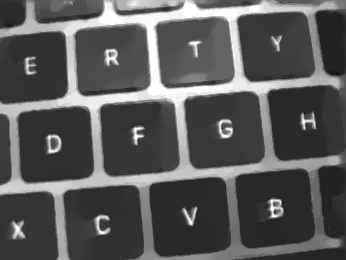}}}
    \\
    \vspace{-6pt}
    \subfloat[LPIPS and SSIM achieved when using the events accumulated over an increasing time period (in milliseconds).]{\includegraphics[width=\exdbevsnuml]{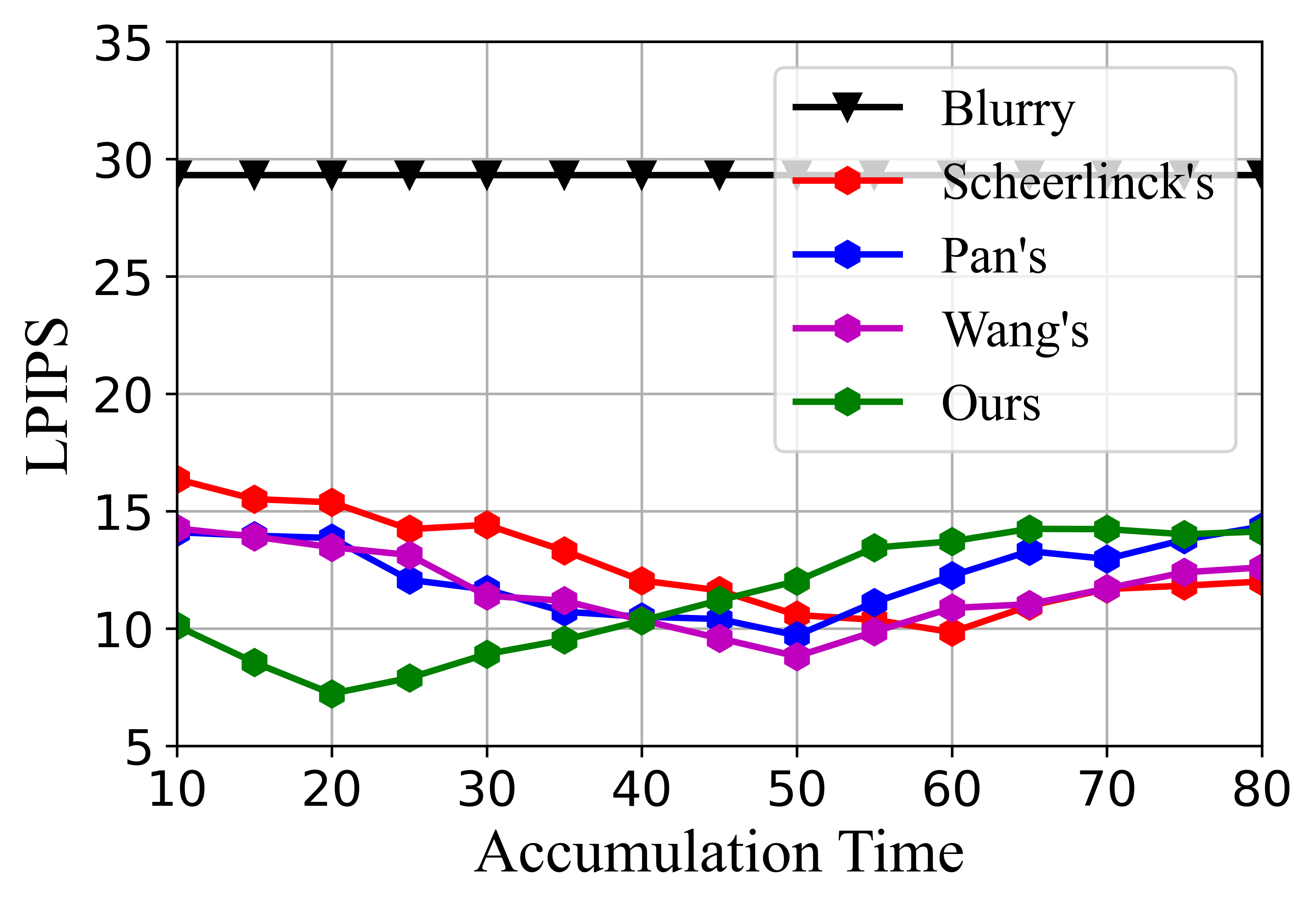}\hspace{-3pt}\includegraphics[width=\exdbevsnuml]{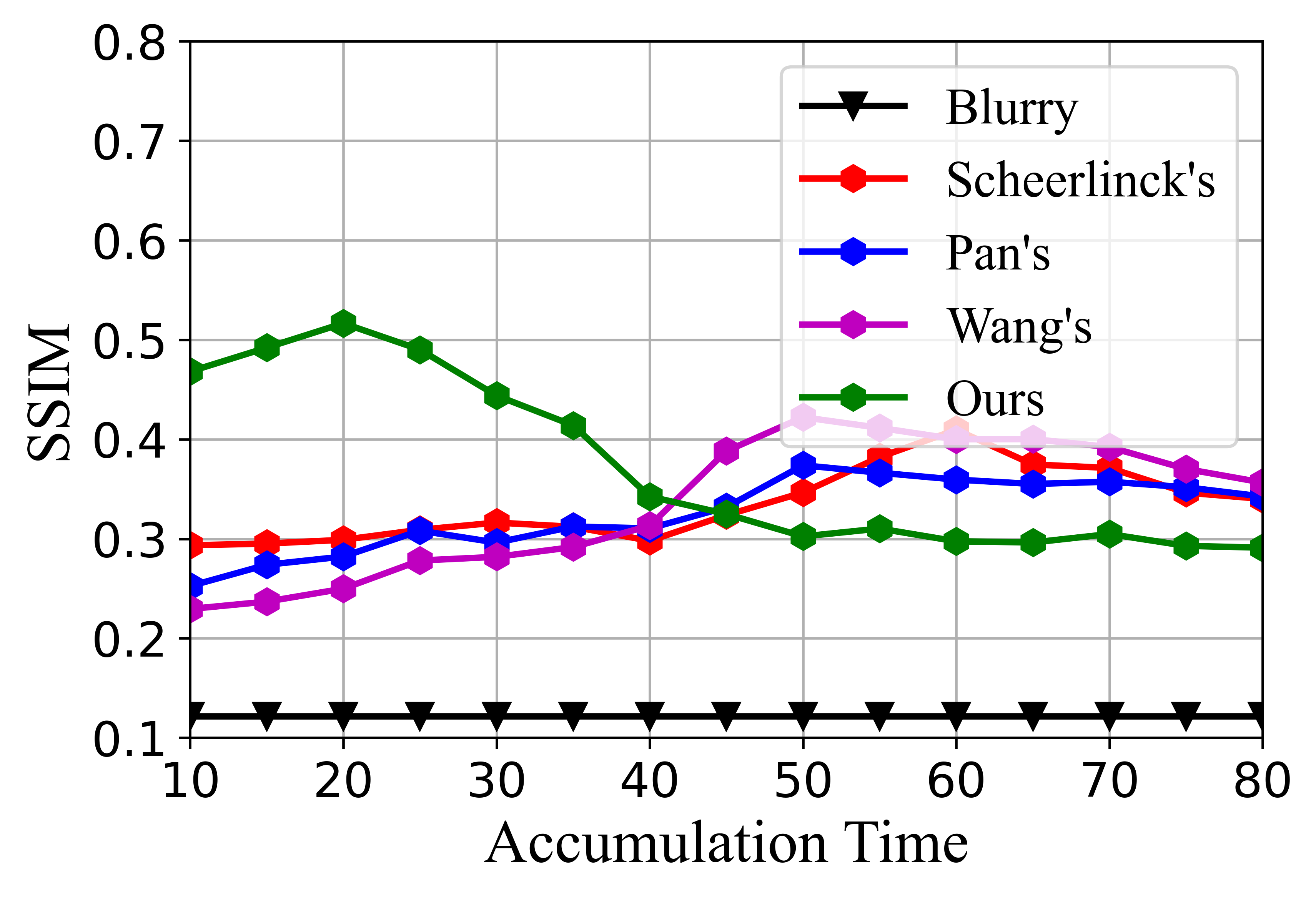}}
    \caption{Visual and quantitative results when exploiting different numbers of events to be complementary information.}
    \label{fig:exdb_evsnum}
\end{figure}

%% file: src/exdb_alpha_beta.tex
\begin{figure}[t]
    \centering
    \subfloat[Blurry Image]{\stackinset{l}{0pt}{t}{0pt}{\includegraphics[width=60pt]{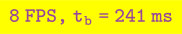}}{\frame{\includegraphics[width=\exdbalphabeta]{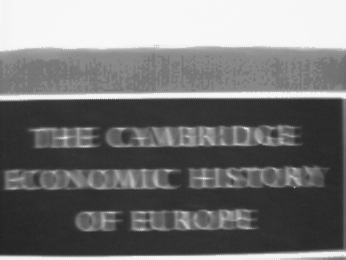}}}}
    \hspace{5pt}%
    \subfloat[$\alpha=0$]{\stackinset{l}{0pt}{t}{0pt}{\frame{\includegraphics[width=32pt]{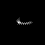}}}{\frame{\includegraphics[width=\exdbalphabeta]{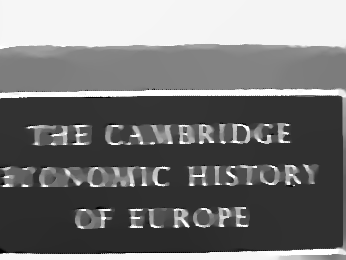}}}}
    \\
    \vspace{-6pt}
    \subfloat[$\alpha=0.24$]{\stackinset{l}{0pt}{t}{0pt}{\frame{\includegraphics[width=32pt]{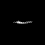}}}{\frame{\includegraphics[width=\exdbalphabeta]{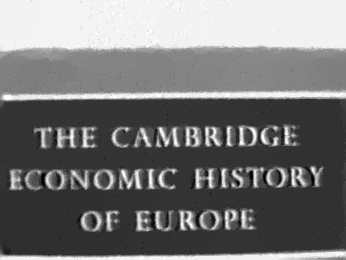}}}}
    \hspace{5pt}%
    \subfloat[$\alpha=0.96$]{\stackinset{l}{0pt}{t}{0pt}{\frame{\includegraphics[width=32pt]{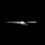}}}{\frame{\includegraphics[width=\exdbalphabeta]{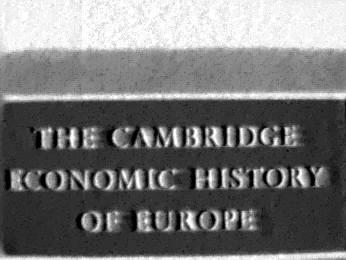}}}}\\
    \vspace{-12pt}
    \subfloat[Effects of the used regularizers on kernel similarity.]{\includegraphics[width=\exdbreg, height=130pt]{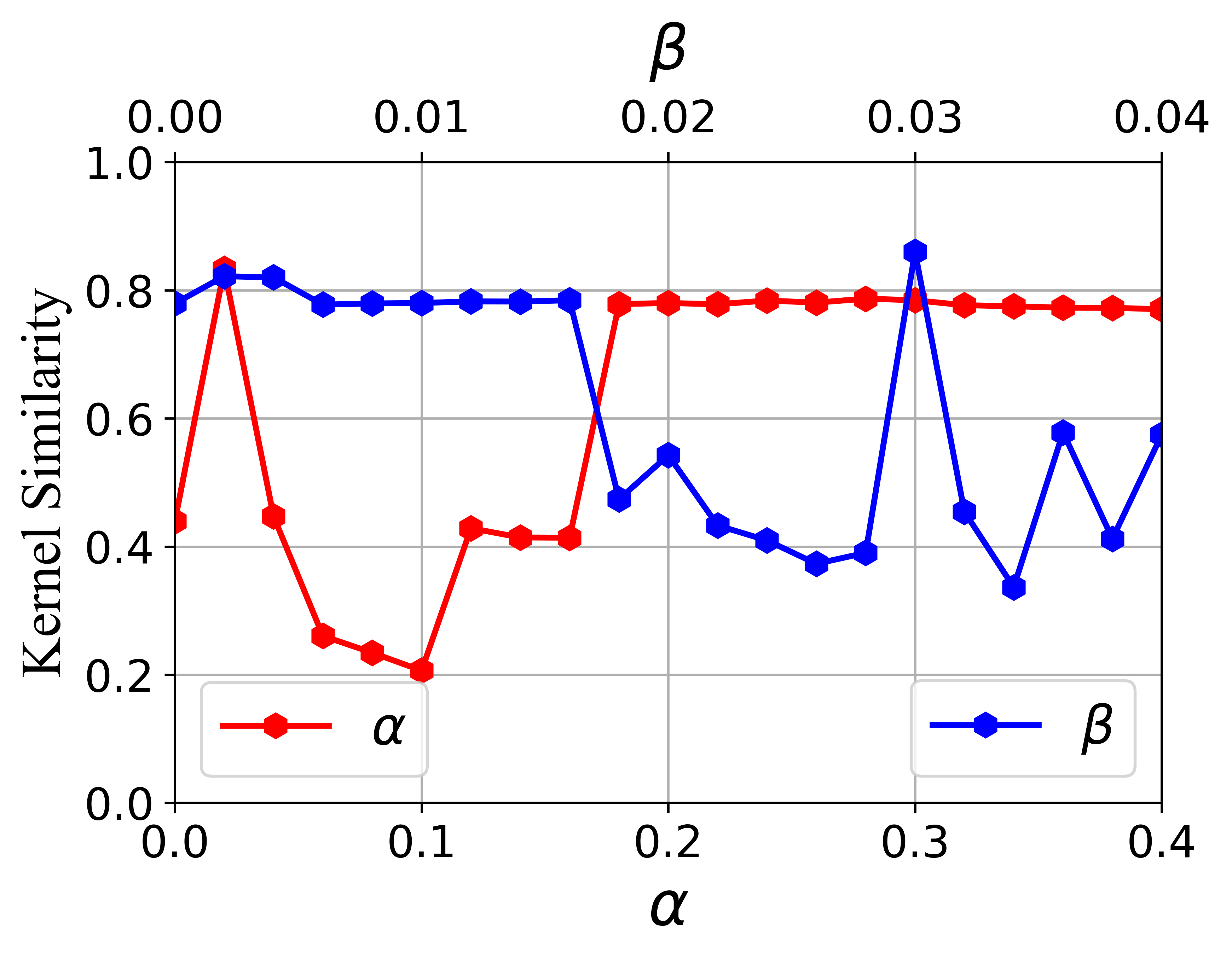}}
    \caption{(a)--(d) Impacts of $\alpha$ (when $\beta = 0.064$) on image deblurring. When $\alpha=0$, the gray area is smoother than that of (a). (c) gives the best result. (d) shows a mass of flecks. (e) Increasing the $\alpha$ value gives high kernel similarity, reflecting that the event-regularized prior precisely describes motion patterns.}
    \label{fig:exdb_alpha_beta}
\end{figure}

%% file: src/prior_quality.tex
\begin{figure}[t]
    \centering
    \subfloat[Blurry Image]{\stackinset{l}{-1pt}{t}{0pt}{\includegraphics[width=60pt]{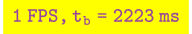}}{\frame{\includegraphics[width=\exdnfactor]{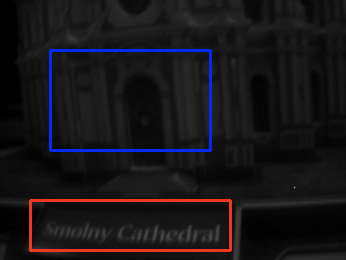}}}}
    \hspace{5pt}%
    \subfloat[$\tau = 2$ \si{\ms}]{\stackinset{r}{0pt}{t}{0pt}{\includegraphics[width=48pt]{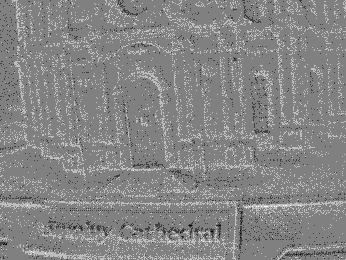}}{\frame{\includegraphics[width=\exdnfactor]{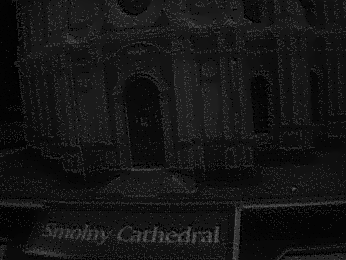}}}}
    \hspace{5pt}%
    \\
    \vspace{-6pt}
    \subfloat[$\tau = 6$ \si{\ms}]{\stackinset{r}{0pt}{t}{0pt}{\includegraphics[width=48pt]{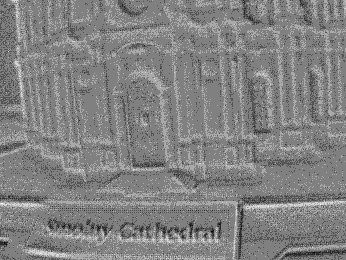}}{\frame{\includegraphics[width=\exdnfactor]{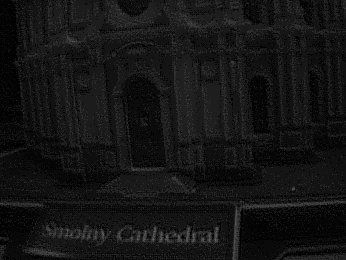}}}}
    \hspace{5pt}%
    \subfloat[$\tau = 16$ \si{\ms}]{\stackinset{r}{0pt}{t}{0pt}{\includegraphics[width=48pt]{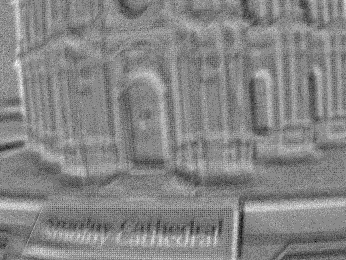}}{\frame{\includegraphics[width=\exdnfactor]{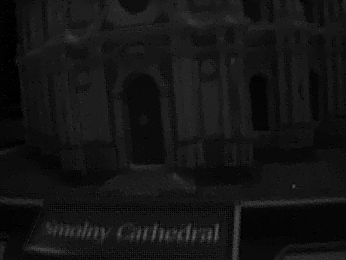}}}}
    \hspace{5pt}%
    \vspace{-6pt}
    \caption{Influences of the quality of the event prior on image deblurring. (a) The blurry image with two areas of concern, and the blue box shows lower contrast. (b)--(d) The deblurred results based on $\mathbf{I}_\tau(t)$ of different $\tau$. The corresponding visualization of $\mathbf{I}_\tau(t)$ is attached.}
    \label{fig:prior_quality}
\end{figure}

%% file: src/exdn_factor.tex
\begin{figure}[t]
    \begin{center}
    \subfloat[Raw Events]{\frame{\includegraphics[width=\exdnfactor]{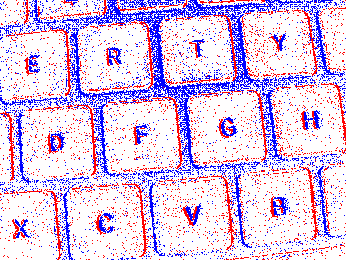}}}
    \hspace{5pt}%
    \subfloat[$\omega = 1$]{\frame{\includegraphics[width=\exdnfactor]{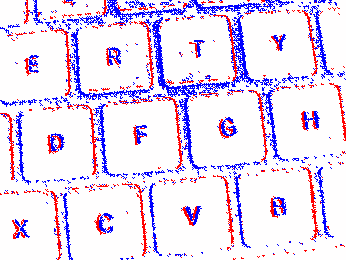}}}
    \hspace{5pt}%
    \\
    \vspace{-6pt}
    \subfloat[$\omega = 10$]{\frame{\includegraphics[width=\exdnfactor]{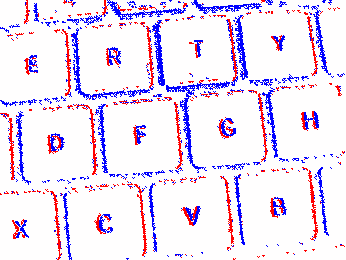}}}
    \hspace{5pt}%
    \subfloat[$\omega = 40$]{\frame{\includegraphics[width=\exdnfactor]{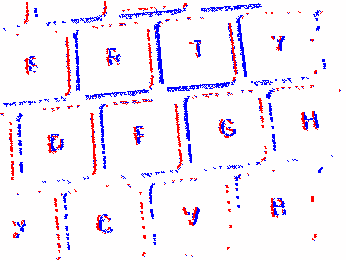}}}
    \vspace{-6pt}
    \end{center}
    \caption{Influences of the calibration factor $\omega$ on denoised events.}
    \label{fig:exdn_factor}
\end{figure}

%% file: src/runtime.tex
\begin{figure}[t]
    \begin{center}
    \subfloat[]{\includegraphics[width=\runtime]{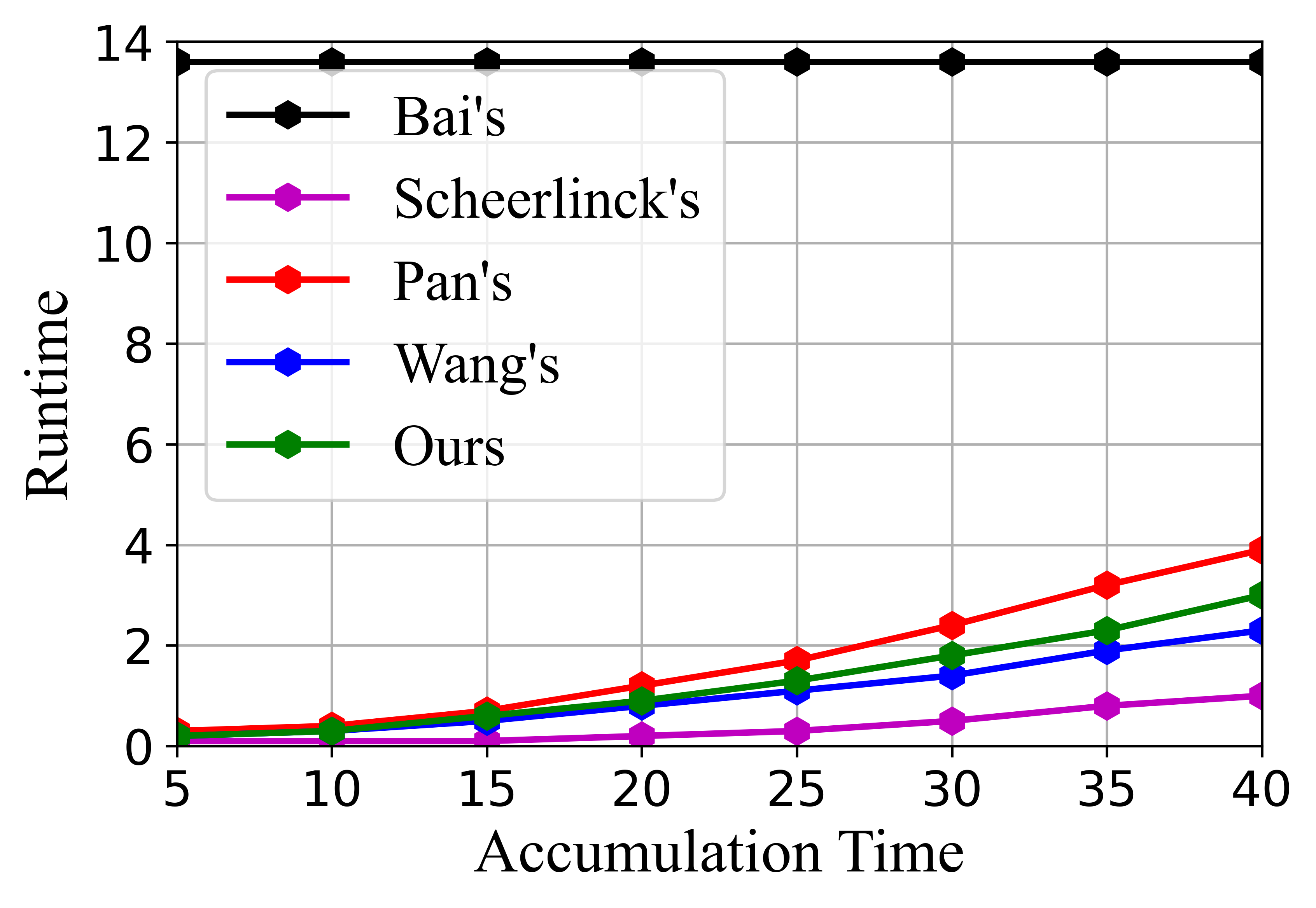}}
    \hspace{-3pt}%
    \subfloat[]{\includegraphics[width=\runtime]{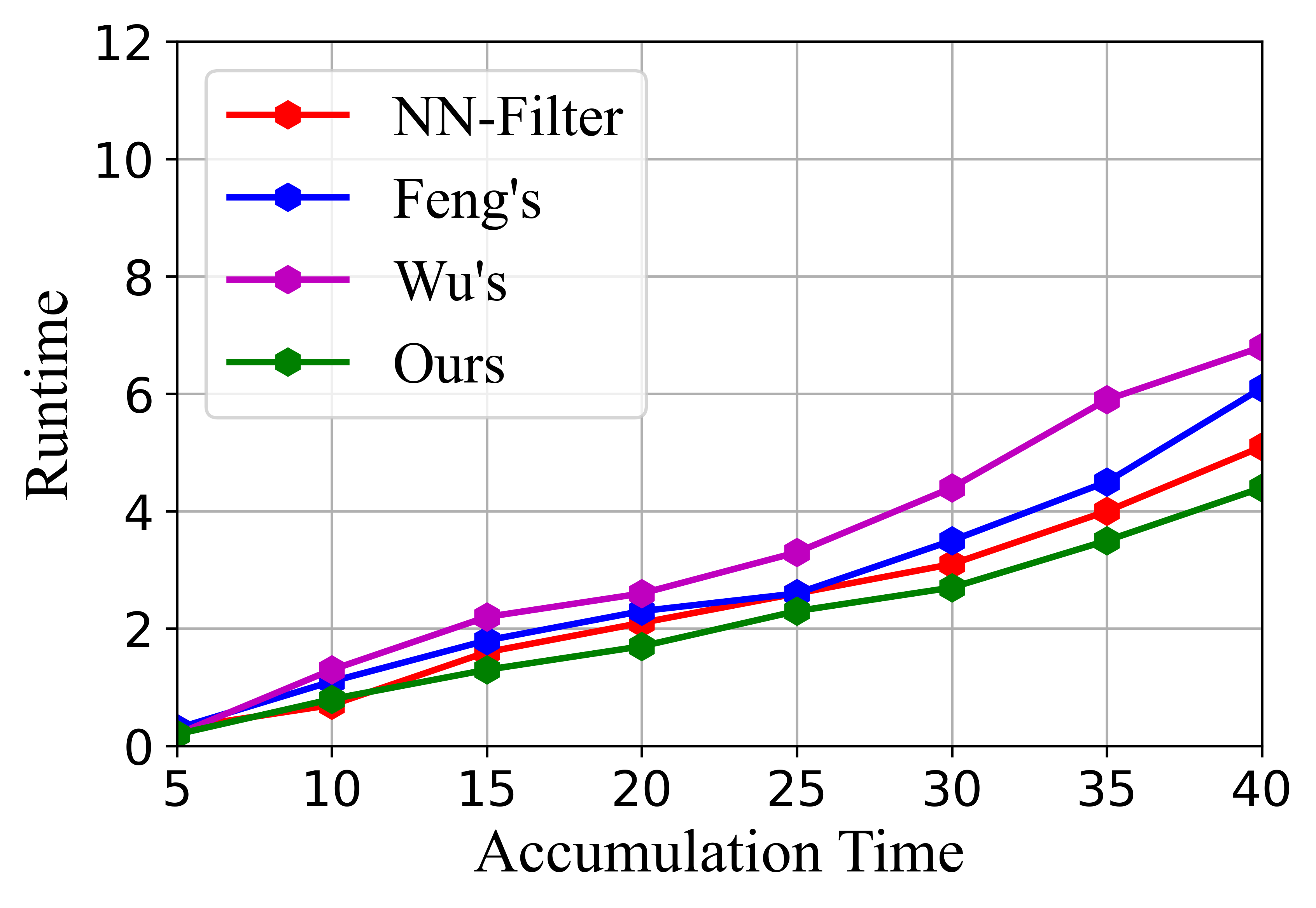}}
    \end{center}
    \vspace{-6pt}
    \caption{With the used events accumulated over an increasing time period (in milliseconds), the runtime (in seconds) required for (a) image deblurring and for (b) event denoising.}
    \label{fig:runtime}
\end{figure}

%% file: src/exdb_nonuniform_app.tex
\begin{figure}[t]
    \begin{center}
    \subfloat[Dong~\etal~\cite{dong2017blind}]{\frame{\includegraphics[width=\exdnfactor]{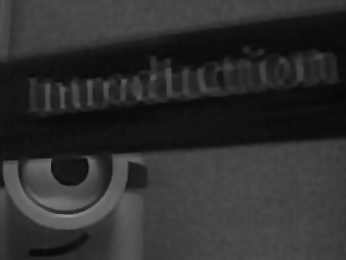}}\hspace{5pt}\frame{\includegraphics[width=\exdnfactor]{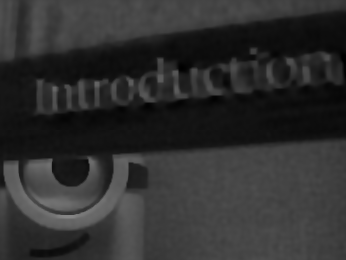}}}
    \hspace{5pt}%
    \\
    \vspace{-6pt}
    \subfloat[Ours]{\frame{\includegraphics[width=\exdnfactor]{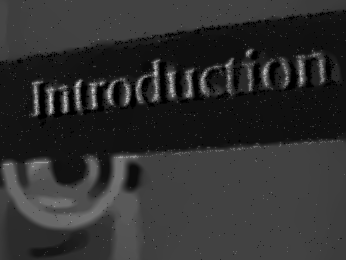}}\hspace{5pt}\frame{\includegraphics[width=\exdnfactor]{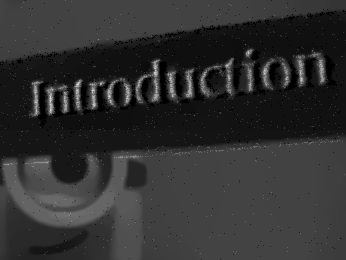}}}
    \vspace{-6pt}
    \end{center}
    \caption{Visual comparisons on deblurring an image with non-uniform blur by processing the image at once (left) or patch-by-patch (right). The examined algorithms keep the same configurations for the two cases.}
    \label{fig:exdb_nonuniform_app}
\end{figure}

%% file: src/dataset_tab.tex
\begin{table*}[t]
\caption{Details of our real neuromorphic dataset.}
\label{tab:dataset}
\centering
\begin{tabular}{l|l|l}
    \toprule
    \textbf{Sample} & \textbf{Setting} & \textbf{Result} \\ \midrule
    \texttt{keyboard} & camera shaking & uniform blurry images; events with noise\\
    \texttt{toy} & camera shaking & uniform blurry images; events with noise\\
    \texttt{text\_cambridge} & camera shaking & uniform blurry images; events with noise\\
    \texttt{cathedral} & camera shaking; weak illumination & uniform blurry images; events with increased noise\\ 
    \texttt{box} & camera shaking; strong illumination & overexposed, uniform blurry images; events with noise\\
    \texttt{text\_number1} & moving object; weak illumination & underexposed, uniform blurry images; events with increased noise\\
    \texttt{text\_number2} & moving object; strong illumination & overexposed, uniform blurry images; events with noise\\
    \texttt{book} & moving object & non-uniform blurry images; events with noise\\
    \texttt{text\_intro} & moving object & non-uniform blurry images; events with noise\\
    \texttt{badminton1} & moving object & non-uniform blurry images; events with noise\\
    \texttt{badminton2} & moving object; strong illumination & overexposed, non-uniform blurry images; events with noise\\
    \texttt{resolution\_fast} & moving object & uniform blurry images; events with noise\\
    \texttt{resolution\_slow} & moving object & uniform blurry images; decreased events\\ 
    \bottomrule
\end{tabular}
\end{table*}

%% file: src/exdb_beta.tex
\begin{figure}[t]
    \centering
    \subfloat[Blurry Image]{\stackinset{l}{-1pt}{t}{0pt}{\includegraphics[width=60pt]{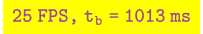}}{\frame{\includegraphics[width=\exdbalphabeta]{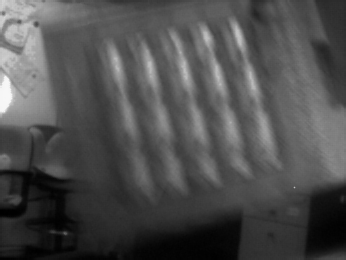}}}}
    \hspace{5pt}%
    \subfloat[$\beta=0.004$]{\stackinset{r}{0pt}{b}{0pt}{\frame{\includegraphics[width=32pt]{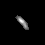}}}{\frame{\includegraphics[width=\exdbalphabeta]{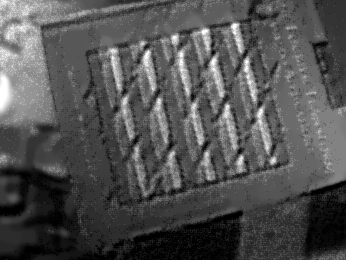}}}}
    \hspace{5pt}%
    \\
    \vspace{-6pt}
    \subfloat[$\beta=0.032$]{\stackinset{r}{0pt}{b}{0pt}{\frame{\includegraphics[width=32pt]{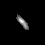}}}{\frame{\includegraphics[width=\exdbalphabeta]{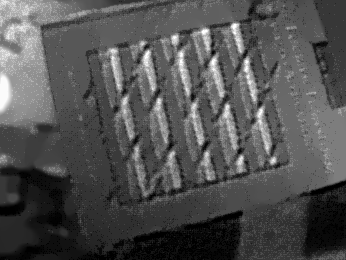}}}}
    \hspace{5pt}%
    \subfloat[$\beta=0.256$]{\stackinset{r}{0pt}{b}{0pt}{\frame{\includegraphics[width=32pt]{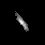}}}{\frame{\includegraphics[width=\exdbalphabeta]{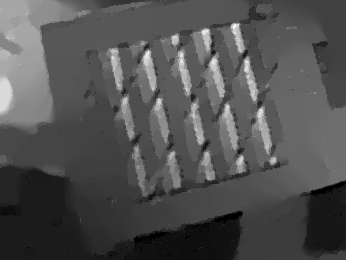}}}}
    \caption{Influences of $\beta$ (when $\alpha = 0.8$) on image deblurring.}
    \label{fig:exdb_beta}
\end{figure}

%% file: src/iterations.tex
\begin{figure*}[t]
    \centering
    \hspace{-9.5pt}%
    \subfloat[Blurry Image and Raw Events]{
    \begin{tabular}[b]{c}
        \stackinset{l}{0pt}{t}{0pt}{\includegraphics[width=50pt]{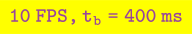}}{\frame{\includegraphics[width=\exiterations]{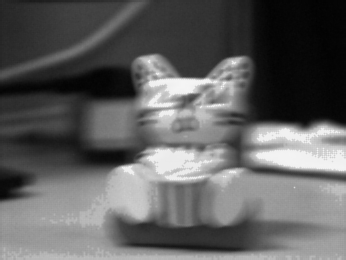}}}\\
        \stackinset{l}{-1pt}{t}{0pt}{\includegraphics[width=70pt]{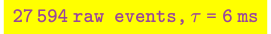}}{\frame{\includegraphics[width=\exiterations]{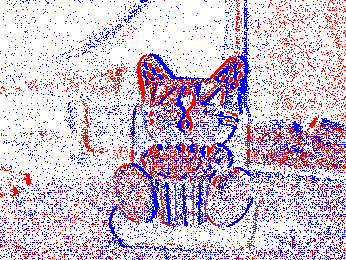}}}\vspace{-2pt}
    \end{tabular}}
    \hspace{-16pt}
    \subfloat[Iteration $1$]{
    \begin{tabular}[b]{c}%
        \stackinset{r}{0pt}{t}{0pt}{\includegraphics[width=24pt]{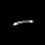}}{\frame{\includegraphics[width=\exiterations]{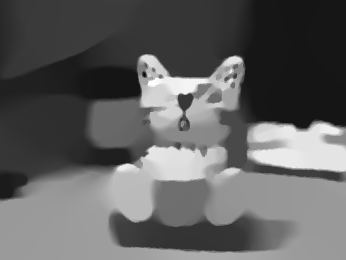}}}\\
        \stackinset{l}{0pt}{t}{0pt}{\includegraphics[width=40pt]{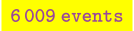}}{\frame{\includegraphics[width=\exiterations]{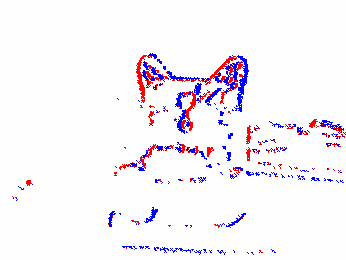}}}\vspace{-2pt}
    \end{tabular}}
    \hspace{-16pt}
    \subfloat[Iteration $2$]{
    \begin{tabular}[b]{c}%
        \stackinset{r}{0pt}{t}{0pt}{\includegraphics[width=24pt]{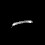}}{\frame{\includegraphics[width=\exiterations]{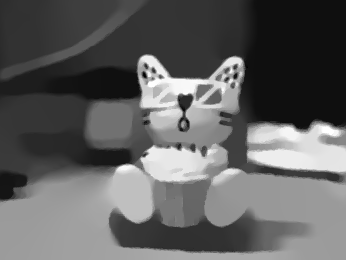}}}\\
        \stackinset{l}{0pt}{t}{0pt}{\includegraphics[width=40pt]{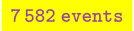}}{\frame{\includegraphics[width=\exiterations]{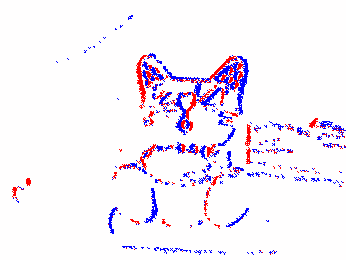}}}\vspace{-2pt}
    \end{tabular}}
    \hspace{-16pt}
    \subfloat[Iteration $3$]{
    \begin{tabular}[b]{c}%
        \stackinset{r}{0pt}{t}{0pt}{\includegraphics[width=24pt]{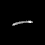}}{\frame{\includegraphics[width=\exiterations]{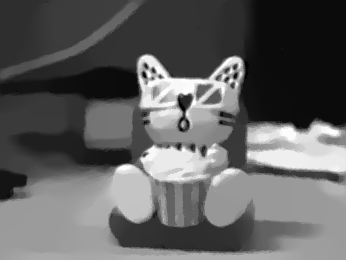}}}\\
        \stackinset{l}{0pt}{t}{0pt}{\includegraphics[width=40pt]{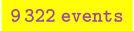}}{\frame{\includegraphics[width=\exiterations]{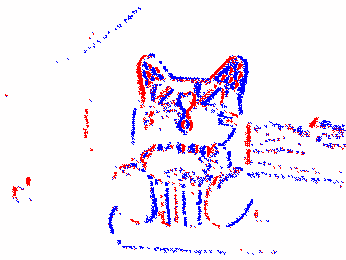}}}\vspace{-2pt}
    \end{tabular}}
    \hspace{-16pt}
    \subfloat[Iteration $4$]{
    \begin{tabular}[b]{c}
        \stackinset{r}{0pt}{t}{0pt}{\includegraphics[width=24pt]{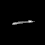}}{\frame{\includegraphics[width=\exiterations]{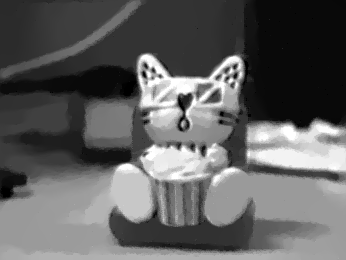}}}\\
        \stackinset{l}{0pt}{t}{0pt}{\includegraphics[width=40pt]{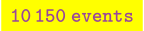}}{\frame{\includegraphics[width=\exiterations]{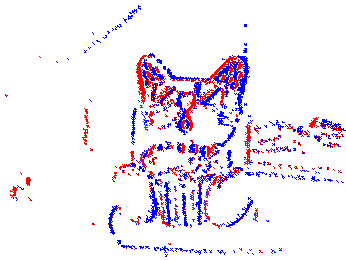}}}\vspace{-2pt}
    \end{tabular}}
    \hspace{-16pt}
    \subfloat[Iteration $5$]{
    \begin{tabular}[b]{c}%
        \stackinset{r}{0pt}{t}{0pt}{\includegraphics[width=24pt]{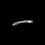}}{\frame{\includegraphics[width=\exiterations]{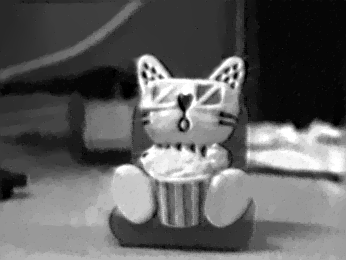}}}\\
        \stackinset{l}{0pt}{t}{0pt}{\includegraphics[width=40pt]{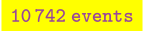}}{\frame{\includegraphics[width=\exiterations]{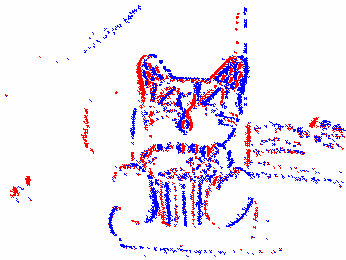}}}\vspace{-2pt}
    \end{tabular}}
    \caption{Visualization of the joint reconstruction of the blur-free image and noise-robust events in each iteration, where the given sample goes through $l_{max}=5$ iterations. The estimated blur kernel and event quantity in a stream are attached.}
    \label{fig:iterations}
\end{figure*}

%% file: src/second_proc.tex
\begin{figure}[t]
    \centering
    \subfloat[The sharp image $\mathbf{S}$ based on the raw events $\mathbf{E}$ with noise. The zoom-in view shows noticeable gray flecks.]{%
        \stackinset{l}{0pt}{t}{0pt}{\frame{\includegraphics[width=32pt]{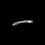}}}{\frame{\includegraphics[width=\secondproc]{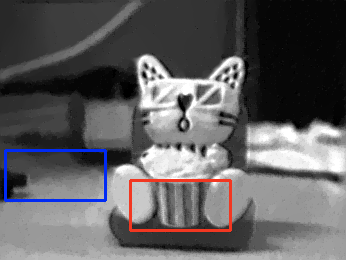}}}%
        \begin{tabular}[b]{c}%
            \frame{\includegraphics[width=\secondprocc]{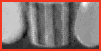}}\\            
            \frame{\includegraphics[width=\secondprocc]{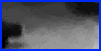}}
        \end{tabular}
    }
    
    \subfloat[The sharp image $\mathbf{S}$ based on the events $\dot{\mathbf{E}}$ with suppressed noise. The zoom-in view shows greatly reduced image artifacts.]{%
        \stackinset{l}{0pt}{t}{0pt}{\frame{\includegraphics[width=32pt]{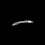}}}{\frame{\includegraphics[width=\secondproc]{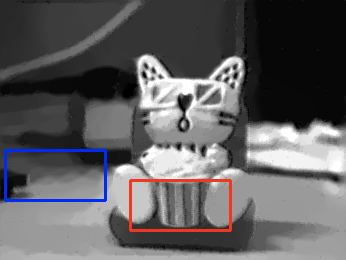}}}%
        \begin{tabular}[b]{c}%
            \frame{\includegraphics[width=\secondprocc]{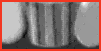}}\\            
            \frame{\includegraphics[width=\secondprocc]{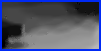}}
        \end{tabular}
    }
    \caption{Investigations into the influence of event noise on deblurred images.}
    \label{fig:second_proc}
\end{figure}

%% file: src/exdb_limitations.tex
\begin{figure}[t]
    \centering
    \subfloat[Ground Truth]{\frame{\includegraphics[width=\exdblimitations]{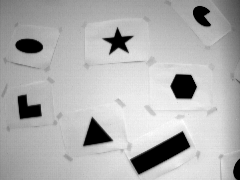}}}
    \hspace{5pt}%
    \subfloat[Synthetic Blurry Image]{\stackinset{r}{0pt}{t}{0pt}{\frame{\includegraphics[width=32pt]{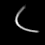}}}{\frame{\includegraphics[width=\exdblimitations]{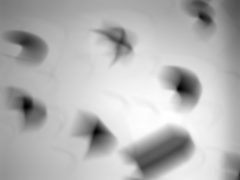}}}}
    \hspace{5pt}%
    \\
    \vspace{-6pt}
    \subfloat[Events]{\frame{\includegraphics[width=\exdblimitations]{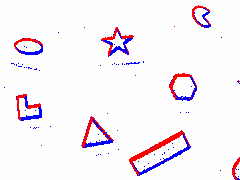}}}
    \hspace{5pt}%
    \subfloat[Our Reconstruction]{\stackinset{r}{0pt}{t}{0pt}{\frame{\includegraphics[width=32pt]{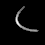}}}{\frame{\includegraphics[width=\exdblimitations]{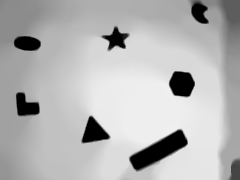}}}}
    \caption{A case where our method loses information during image deblurring.}
    \label{fig:exdb_limitations}
\end{figure}